\documentclass[amsmath,amssymb,twocolumn]{revtex4}
\usepackage{mathrsfs}
\usepackage{graphicx}  
\usepackage{color}
\usepackage{cases}
\usepackage{array}
\newcommand{\dbar}{{\mathchar'26\mkern-11mud}}

\begin{document}

\title{Comprehending robust quantum effects in organic semiconductors: Charge-transfer excitons, aggregate phonons and fractons}

\author{Yao Yao$^{1}$\footnote{Electronic address:~\url{yaoyao2016@scut.edu.cn}}}

\address{
$^1$Department of Physics and State Key Laboratory of Luminescent Materials and Devices, South China University of Technology, Guangzhou 510640, China}

\date{\today}

\begin{abstract}
In organic semiconductors working under ambient circumstance, there are remarkable quantum effects lack of comprehensive understanding, and an exotic composite particle named charge transfer (CT) exciton is normally regarded as the key ingredient. Another essential substance is the phonons stemming from intra- and inter-molecular vibrations, and the relevant diagonal electron-phonon couplings give rise to spatial localization and the off-diagonal couplings refer to the dispersion of electron wavefunctions. In this work, we first propose a toy model based upon the fracton physics to phenomenologically unveil the coherent motion of CT excitons followed by resonant phonons in the aggregates. Based on this model, a generic scenario of hierarchical quantum effects is described in various material systems. To examine whether the fracton model can be applicable in practice, we calculate the out-of-time-ordered correlator (OTOC), a quantum dynamic measurement of the entanglement entropy, of CT excitons by the adaptive time-dependent density matrix renormalization group algorithm. On the basis of three commonly-used realistic models in organic materials with two competing interactions taking into account, we investigate the irreducible roles of CT excitons and aggregate phonons in the ultrafast charge transfer, the coherent polaron hopping, and the dissociation of triplet pairs, respectively. Our theory unifies the charge, exciton, spin and phonons into a single framework, which may help clarify the complicated and diverse quantum effects in organic semiconductors.
\end{abstract}

\maketitle

\section{Introduction}

Organic semiconductors are increasingly attractive in the field of photoelectricity, as the organic light emitting diodes have been successfully industrialized and the organic photovoltaics and spintronics are of high potential in applications \cite{Review1}. Thanks mainly to the cost containment, the single crystal phase of organic materials is rarely present in practice and the utilization of molecular aggregates are always in fashion up to date \cite{Review2}. The ubiquitous disorders induced by the local thermal vibrations of molecules (phonons) intuitively give rise to the localization of every eigen-states of electrons with the spatial localization length being around one molecule \cite{Local1,Local2}. On the other hand, in presence of the strong nonlocal electron-phonon (vibronic) couplings or the so-called off-diagonal dynamic disorders, it is assumed that the charge transport can be bandlike or quantum coherent implying the electron is in some sense delocalized \cite{Troisi1,Troisi2}. The delocalization is also regarded to be essential in other robust quantum effects such as the coherent charge separation and ultrafast hole transfer in photovoltaics \cite{Delo1,Delo2,Delo3,Delo4,Delo5,Delo6,Delo7,Delo8,Yao19}. This localization-delocalization duality of electrons in organic semiconductors long-termly serves as the major puzzle for the community \cite{Review3}, and in this context the thermodynamics and statistical mechanics in organic semiconductors have to be reexamined.

Before proceeding, let us first clarify the terminology. In terms of eigenstate thermalization hypothesis (ETH) \cite{MBL0}, the spatial localization implies the memory of initial condition could be preserved in local integrals of motion, so the dissipation and thermalization are absent suggesting a complete localization serves as a pure quantum effect. The delocalization then correspondingly refers to the relaxation of wavepacket in the ergodic or thermalized phase, which can be ascribed to classical physics. On the contrary, however, by the common understanding on organic semiconductors, the localized charge and exciton are regarded as (semi-)classical particles and the quantumness stems from wavefunction spreading on spatial dimension which in chemical language is nothing but the delocalization. This terminological contradiction reflects the different angle of observer's view for quantum effect and different custom from different fields, as by definition, the quantum coherence reflects the correlation between different quantum states, which strongly depends on the choice of preferred basis of observers \cite{Quantum}. There are a huge number of electrons in realistic materials. In condensed matter physics, people in most cases care about the doped electrons and consider the intrinsic ones as background mean-field potential. The primary source of dissipation for these doped electrons turns out to be the scattering at impurities or defects so that the coherence is mainly with regard to the momentum space. In molecules, however, the view angle focuses on the real space. This is because the conducting electrons, no matter from closed or open shells, are normally induced by molecular vibrations via topological polaron effect which may concurrently give rise to decoherence. As a result, the terminology becomes quite controversial, since we indeed want localization of electrons to produce robust quantum effects but at the same time there are numerous localized electrons in molecules we do not like. So let us be more explicit. We do not prefer the localization stemming from the negative effects of phonons (see discussions below); our motivation is to search for other more robust ingredients to induce the localization and thus the macroscopic spontaneous symmetry breaking, such as topology, dimensionality, two-body interactions and so on, similar with that effort of Su, Schrieffer and Heeger for topological elementary excitations in organic metals \cite{SSH}.

In the traditional theory of insulator, the Anderson localization plays the essential role in the electron transport \cite{Anderson}. With relatively weak localization, Mott's variable-range hopping mechanism matters \cite{Mott}. In terms of delocalized Bloch waves, the band theory achieved great successes. All these theories are more or less correlated to the thermodynamics and statistical mechanics of (near) equilibriums. The picture of free electron gas is normally applicable, and a single term such as the static disorder or the single electron hopping in the lattice dominates the transport mechanisms. The traditional techniques of transport measurement were thus established on the basis of these widely-accepted theories. In presence of strong many-body interactions, however, the scenario could be much more complicated. For example, given numerous local conserved quantities in a spatially translational invariant system, there exist some novel elementary excitations such as anyons and fractons which can freely move in certain dimension without dissipation \cite{Toric,Fracton0,Fracton1,Fracton2,Fracton3,Fracton4}. In addition, in the strong vibronic coupling systems, the thermal equilibrium is not easy to be achieved, and the dynamic processes out of equilibration become important. In this context, when various many-body interactions such as Coulomb interactions and vibronic couplings are concurrently considered in, e.g., charge-transfer (CT) states, there should be not an apparent small parameters to make the perturbation theory available, and the competition among different terms results in complicated many-body effects \cite{Review3}. As a consequence, the quantum effect emerges in such organic semiconductors with strong interactions and it is always necessary to develop a quantum dynamics theory to bridge the microscopic mechanisms and the device measurements.

In those complicated problems, the central point in our opinion is to uncover a mechanism that the quantum coherence survives the warm and wet environment. In particular, the ubiquitous thermal vibrations must destroy the quantumness in an easy manner; it must be protected by a sufficiently robust mechanism as stated. A recent work by Sous and Pretko mentioned a novel quasi-particle, named ``fracton", in the electron-phonon coupling systems \cite{Fracton0}. The fracton model was first established in a so-called X-cube structure for majorana fermions \cite{Fracton1}. Soon after that, researchers from various fields indicated very appealing characteristics of fractons \cite{Fracton2}. According to the rank-two tensor gauge theory, the fracton can be regarded as an electric dipole moving in limited dimensions \cite{Fracton3,Fracton4}. The work of Sous and Pretko then suggested that the motion of a single polaron will generate phonons and become immobile up to the sixth order, while a pair of two polarons can freely move up to the second order perturbation. The key point of them is that, when two polarons go together, the behind one can instantly absorb the phonons that the ahead polaron releases. This scenario greatly enlightens us to understand the quantum effects in organic semiconductors in an alternative view angle, that is, a CT state formed by two polarons might be the essential issue closely linked to the quantum effects.

On the experimental side, in order to distinguish the quantumness in charge transport, one can measure the mobility-temperature ($\mu-T$) relationship in a macroscopic manner and assign the positive and negative dependency to localized and delocalized phase, respectively \cite{Review4,Review5}. With the novel findings of ultrafast spectroscopy in bloom, studies on the dynamics of quantum coherence in organic semiconductors emerged to be prevalent in recent years \cite{Cohe1,Cohe2,Delo1,Delo2,Delo3,Delo4,Yao15,Yao16}. The mobility measurement commonly refers to the real space while the ultrafast spectra account for the excitations and emissions in the energy space, making the assignment of quantum coherence complicated. Moreover, nonspecialists are always easy to be confused with the electron coherence, the exciton coherence, the spin coherence and the vibronic coherence. To this end, a quantity that is basis-free should be remarkably useful for getting rid of the confusion, and the entropy turns out to be a good candidate \cite{Entropy1,Entropy2,Entropy3,Entropy4,Entropy5,Entropy6,Entropy7,Entropy8,Entropy9,Entropy10,Entropy11,Entropy12,Entropy13}. However, the state-of-the-art experiments mostly focus on the entropy in a statistical manner, that is, the number of diabatic states or molecular frontier orbitals, irrelevant to the quantum coherence \cite{Entropy11}. The quantum-mechanical entropy, namely the quantum entanglement, characterizes the correlation among substances, and using the entanglement entropy as a fingerprint in organic semiconductors can largely help comprehend the complicated and diverse quantum effects related to the charge, exciton, spin and phonon dynamics. Furthermore, people employ the ultrafast spectroscopy to reveal the microscopic processes with the timescale being femtosecond to picosecond, while the devices work in a macroscopic circumstance. The entropy should also play essential roles in bridging the microscopic mechanisms and macroscopic effects, which also serves as an essential motivation of the present paper.

The paper is organized as follows. In the next Section, we first describe a generic scenario of entanglement entropy in terms of quantum heat engine model, and then propose a toy model for describing the adiabatic motion of CT excitons followed with the discussion of multi-molecular resonant vibrational modes (we call them ``aggregate phonons"). In the following Section, we discuss the hierarchy of quantum effects in the realistic materials and then typical kinds of organic molecules are briefly introduced. In Section IV we first introduce a quantity to measure the quantum entanglement of CT states which is named out-of-time-order correlator (OTOC), and then discuss the coherent exciton and CT state dynamics stemming from the competition between the Coulomb interaction and nonlocal vibronic coupling. The charge transport which mainly refers to the competition between diagonal and off-diagonal vibronic couplings, and the singlet fission with the competition of spin-spin interaction and the local bosonic environment are explored as well. The last Section is for the summary and outlook.

\section{Framework}

Our main scope in this work is to explore a mechanism that the quantum coherence may survive the warm and wet environment in which the realistic organic semiconducting devices are working. This is not to say we are proposing a scenario to observe macroscopic quantum effects wherein such as superconducting and quantum Hall effect, although there is great chance for their appearance. In amorphous organic aggregates, the size and degree of disorders of the micro-structure varies within a large extent, so it is intuitive that the classical and quantum effects should coexist, namely some micro-structures are classical and others are quantum. We are herein focusing on the latter.

\subsection{Entropy}

Let us begin with thermodynamics. In order to maintain the quantum coherence alive, the thermodynamic entropy must not be increased according to the second law, which means by definition the conducted mechanisms below must be \textit{adiabatic}. We herein borrow the language of quantum heat engine model to make the discussion of entropy smoothly connecting to the quantum dynamics. The first law of thermodynamics tells us that \cite{Statistic}
\begin{eqnarray}\label{dUt}
dU=\dbar Q+\dbar W,
\end{eqnarray}
where $U$ is the internal energy, $Q$ is the heat and $W$ is the work. For a system with the Hamiltonian being $\hat{H}$, the internal energy can be written as
\begin{eqnarray}
U=\sum_np_n\varepsilon_n,
\end{eqnarray}
where $\varepsilon_n$ is the energy of $n$-th eigenstate of $\hat{H}$ and $p_n$ is the relevant population. Taking the differential of $U$, we obtain
\begin{eqnarray}\label{dUq}
dU=\sum_n\left(\varepsilon_ndp_n+p_nd\varepsilon_n\right).
\end{eqnarray}
If the system is initially in a canonical ensemble, $p_n=\exp(-\varepsilon_n/k_{\rm B}T)/Z$ with $Z$ being the partition function and $T$ being the temperature, the first term on the right hand side of Eq.~(\ref{dUq}) turns out to be
\begin{eqnarray}
\sum_n\varepsilon_ndp_n=-k_{\rm B}T\sum_nd(p_n\ln p_n).
\end{eqnarray}
With the definition of entropy in the canonical ensemble being $S=-k_{\rm B}\sum_np_n\ln p_n$, this term is nothing but $\dbar Q=TdS$. As a result, the differential of work is then given by $\dbar W=\sum_np_nd\varepsilon_n$, implying the change of the eigen-energies refers to the work performed by a generalized force conjugated to the generalized coordinates \cite{QHE1,QHE2,QHE3,QHE4,QHE5}. Therefore, the thermodynamic quantities in the statistical mechanics, the heat and the work, are related to the quantum dynamic quantities, the change of population and eigen-energy, respectively. We are thus able to analyze the macroscopic thermodynamics of organic semiconductors in the microscopic scenario of the quantum heat engine \cite{QHEO1,QHEO2,QHEO3,QHEO4,QHEO5}.

In the microscopic scope, the change of the eigen-energy spectrum of the Hamiltonian stems from the explicit temporal dependence of the Hamiltonian. Considering the Berry phase is not detectable in a macroscopic manner in disordered molecular materials \cite{WeakLocal}, when the Hamiltonian changes with time sufficiently slowly, the population distributions $\{p_n\}$ on the eigen-energy spectrum do not change so we can call this process as adiabatic or coherent ($\dbar Q=0$) \cite{QHE4}. The instantaneous eigen-states of the Hamiltonian can be named as the adiabatic states. Subsequently, the heat exchange is regarded to be originated from the nonadiabatic and incoherent hopping among the adiabatic states ($\dbar W=0$).

In organic semiconductors, what makes the Hamiltonian of electrons time dependent is the vibronic coupling to nuclei which normally move slower than electrons. The diagonal (intramolecular) vibronic coupling serves as the local energy disorders to induce the spatial localization. In terms of the adiabatic approximation, the diagonal couplings do not drive the electrons move, so the local disorders can be regarded as ``static" in some sense. Subsequently, the diagonal couplings do not prefer spatial quantum coherence of electrons. On the other hand, the off-diagonal (intermolecular) vibronic coupling and thus the dynamic nonlocal disorder enables the nonadiabatic hopping (i.e. the conical intersection) between nuclear potential surfaces and also changes the heat and entropy \cite{CI1,CI2}. This refers to nonadiabatic processes so one can expect the quantum coherence will be quickly lost. As a result, no matter which case, phonons are always a negative factor for the quantum effects.

One might then be doubting that, there are some spectra of phonon modes that may induce non-Markovian effect so the phonons are not always negative. This viewpoint long-termly serves as the essential motivation of the researches of quantum coherence in this system. In our previous works it has been discussed that \cite{Yao15,Yao16}, through a unitary transformation, the independent vibrational modes are mapped onto a half-infinite bosonic lattice with disordered bosonic on-site energy and hopping terms. The memory of initial state could be reserved at somewhere on the bosonic lattice and in a certain duration the quantum entanglement revives. In a heat engine viewpoint, this rebirth of quantum coherence is indeed able to enhance the efficiency of macroscopic devices. But in a dynamic perspective, this effect can almost not be observed due to the fluctuation of decoherence time.

We then notice that, in presence of both disorders and many-body interactions, there may emerge a novel effect which is so-called many-body localization (MBL) \cite{MBL0,MBL1,MBL2,MBL3}. In the MBL phase, the electronic eigen-states are localized, but the quantum information can be transferred in a logarithmic manner since the entanglement entropy (not thermodynamic entropy) is persistently and slowly improved \cite{MBL4,MBL5}. This is because the number of many-body states is overwhelmingly larger than that of single-body states and the many-body interaction enables the transition between localized many-body eigen-states. If some kind of MBL effects is actually present in organic semiconductors with strong interactions, the novel phenomena relating to the quantum coherence and insensitive to the thermal energy, such as the ultrafast hole transfer in nonfullerene cells, can be well understood \cite{Yao19}. Rather, the MBL effect does normally not survive the ambient circumstance due to the ``domino" effect of thermal phase \cite{domino}. So even if the MBL does exist, it should be with respect to the rare region effect. With the state-of-the-art technique of synthesis, one can observe the dynamic effect at most on the mesoscopic level, which is then worth discussing. Macroscopic MBL bulk can not be expected so far.

\subsection{CT states}

\begin{figure}
\includegraphics[scale=0.45]{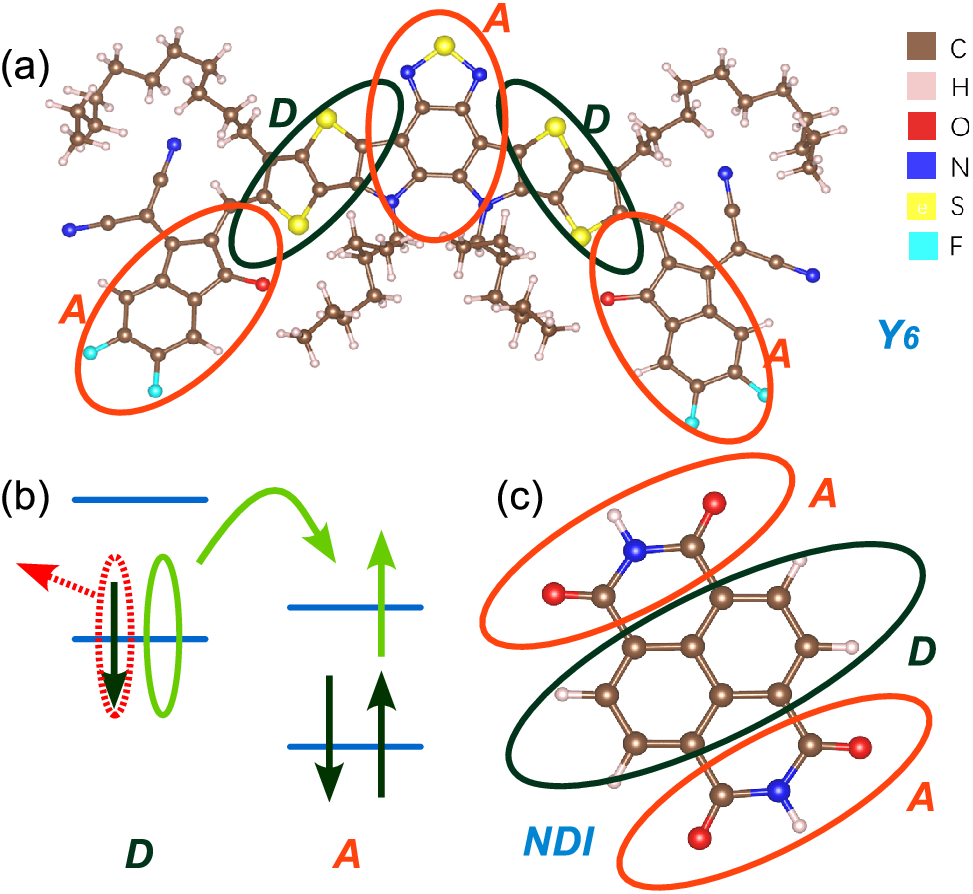}
\caption{(a) The chemical structure of Y6 with the D and A units highlighting. (b) On the D-A structure when an electron on D's HOMO is transferred to the neighboring A's LUMO, the remaining electron will try to escape to lower the chemical activity. (c) The chemical structure of NDI molecule.}\label{Fig1}
\end{figure}

To explore the many-body effect, we first consider the CT state in which electron and hole strongly interact with each other by Coulomb attraction. It is a practical experience for material synthesis that a successful acceptor molecule with high performance should possess well-designed donor-acceptor (D-A) hetero-structure. For example, the celebrating nonfullerene small molecule Y6 possesses a complicated A-D-A-D-A structure as sketched in Fig.~\ref{Fig1}(a) \cite{Y6}. On the other hand, the cells are by default fabricated with bulk heterojunction structure of donor and acceptor molecules such as PM6/Y6 system. Nonspecialists might be confused with this nomenclature. To simplify our discussion, let us merely consider the film phase without regard for solution phase with individual molecules. Then the D and A units denote the moieties which can be on the backbones, side or end groups. As our previous work has indicated \cite{Yao18}, due to the attractive interaction the D and A units naturally form an alternating structure and the ground state of this structure turns out to be staggered positive and negative charge densities which was named ``self-accumulated intrinsic charges". In the present terminology, this is a liquid of CT states.

Researchers from fields of traditional semiconductors would image the CT states as pairs of free electron and hole, so that it can be modeled by two fermion operators. This picture would be easily confused with that of the separated electron and hole appearing in the end stage of exciton dissociation, in which the quantum coherence or entanglement between the electron and hole is completely lost. To this end, we first have to define a set of operators specified for CT states.

Let us consider a pair of D and A units, each of which possesses a highest occupied molecular orbital (HOMO) and a lowest unoccupied molecular orbital (LUMO), as sketched in Fig.~\ref{Fig1}(b). If we consider both of them are with closed shell, the HOMOs are initially occupied by a singlet pair of electrons while the LUMOs are empty, which is the neutral state. If open shells are taken into account, the singlet could be replaced by doublet for radicals and triplet for diradicals, which will be discussed later on. It is then easy to see that there are two states for this D-A struture: One is the neutral state and the other is the CT state, namely one electron transfers from D's HOMO to A's LUMO. Which state has lower energy depends on the relative energy of these two orbitals. In well-optimized D-A structures, they are near-degenerate so in the ground state of the system there should be great contribution from CT state. It is also worth noting that, charge doping is not a regular method in organic semiconductors, so we have not to consider the cases that the total electron number is not equal to four.

Since two states are now in play, we can define a set of Pauli-like operators $\hat{X}$ and $\hat{Z}$ to denote the off-diagonal and diagonal operations on the CT states. That is,
\begin{eqnarray}\label{ZX}
\hat{X}_{\sigma}=\hat{c}^{\dag}_{\rm A,\sigma}\hat{c}_{\rm D,\sigma}+\hat{c}^{\dag}_{\rm D,\sigma}\hat{c}_{\rm A,\sigma},~~~\hat{Z}_{\sigma}=\hat{n}_{\rm A,\sigma}-\hat{n}_{\rm D,\sigma},
\end{eqnarray}
where
$\hat{c}^{\dag}_{\rm \mu,\sigma}$ ($\hat{c}_{\rm \mu,\sigma}$) creates (annihilates) an electron on D's HOMO or A's LUMO with spin $\sigma$, and $\hat{n}_{\rm \mu,\sigma}$ is the relevant operator of electron number. Similar with the usual Pauli operators, $\hat{X}^2=\hat{Z}^2=1$ and $\{\hat{X},\hat{Z}\}=0$. In absence of the external magnetic field, the physical spin $\sigma$ is not important, so we can neglect it for simplicity.

The operator $\hat{Z}$ has two eigenstates: the neutral state with eigenvalue $-1$ and the CT state with eigenvalue $+1$. In normal cases, the energy of CT state is slightly higher than that of neutral state. Similar with the usual hopping term of electron, we can act $\hat{X}$ on different D-A bonds to enable the movement of CT states. Since $\hat{X}$ and $\hat{Z}$ do not commute with each other, the motion of a single CT state has to exchange the energy with environment and can not be adiabatic, so it is an entropy-increasing process. This implies the quantum coherence for a single CT state can only be alive within a very short lifetime and small spatial extent, which is the same with that of a single polaron \cite{Fracton0}. Consequently, a single CT state does not serve as a robust protection for the quantumness, so we have to look for more profound mechanisms.

\subsection{Toric code model and CT excitons}

\begin{figure}
\includegraphics[scale=0.45]{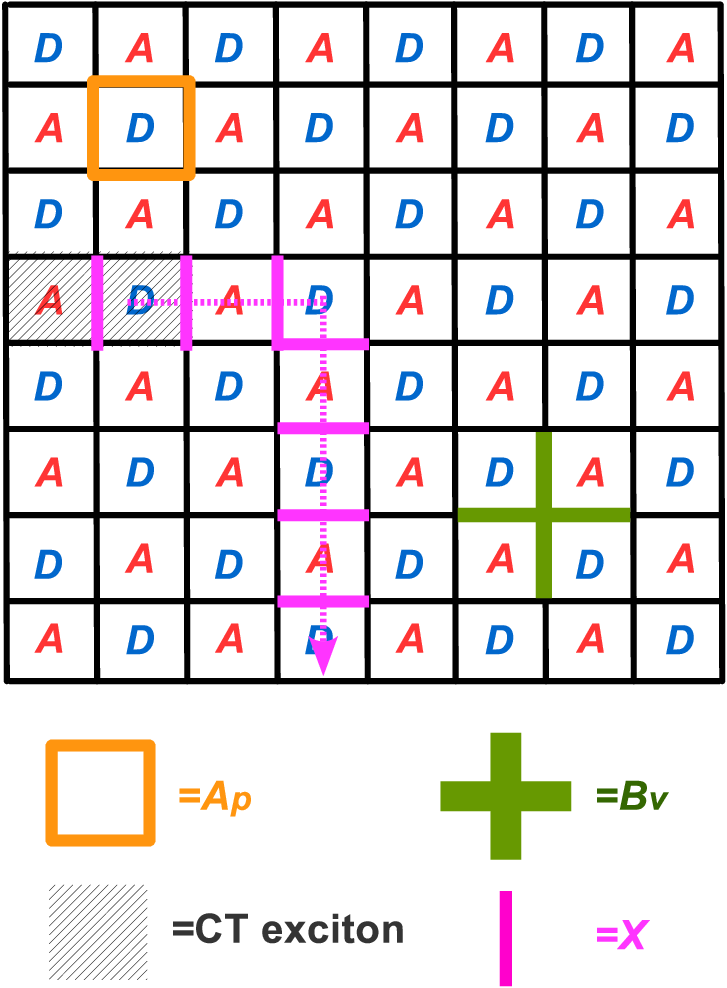}
\caption{Toric code model for CT excitons. The 2D lattice is formed with alternating D-A structure. $\hat{A}_p$ and $\hat{B}_v$ operators act on plaquette and vertex, respectively. When a D-A bond (link) is flipped by an $\hat{X}$ operator, two excited states are generated represented by the shaded area. The arrow denotes the moving direction of a CT exciton by continuously acting $\hat{X}$ operator on relevant link.}\label{Fig2}
\end{figure}

If we consider the CT state as an electric dipole, one can intuitively connect it with the rank-two tensor gauge theory, or more recently the fracton physics, which accounts for the motion of dipoles in limited dimensions \cite{Fracton3,Fracton4}. The starting point of fractons is the Kitaev's toric code model for anyons \cite{Toric}. Our following discussion is strongly motivated by this celebrated model.

We noticed the toric code since realizing the following scenario. As sketched in Fig.~\ref{Fig1}(b), when an electron transfers from the D's HOMO to the neighbouring A's LUMO, the remaining electron on the D's HOMO becomes a free radical with high chemical activity and would prefer to transfer to another A's LUMO to lower the instability. It means in this situation a single CT state possesses higher free energy than two pairing CT states, which is quite similar with the relationship between CO and CO$_2$. As an instance let us consider the naphthalene diimide (NDI) molecule which holds a typical A-D-A structure with the backbone naphthalene being D unit and two imide end groups as A units, as sketched in Fig.~\ref{Fig1}(c). Experience tells us that, if one donates an electron to one of the imide groups to generate a single radical, it is natural that the other imide will want to withdraw an additional electron. As a result, there appears a mixture of radical and diradical in an NDI molecule.

Let us thus consider a more specific case, i.e., D and A units form an alternating structure on a two-dimensional (2D) lattice, as displayed in Fig.~\ref{Fig2}. This is an extended model of that in our previous work \cite{Yao18}. We note here that, although the schematic diagram seems as a toy model, it is straightforward to be substantiated in realistic materials. For example, for the NDI molecules with linear A-D-A structure, they will form a 2D lattice through self-assemble by producing the $\pi-\pi$ stacking on the normal direction and hydrogen bonds between the imide end groups. Another example is the new kinds of nonfullerene solar cells, such as PM6/Y6. The A-D-A-D-A structure of Y6 naturally has a twist angle which convincingly gives rise to the structure of the plaquette. We then make a statement that, as long as some kind of quantum effect survives within a certain spatial extent much larger than one molecule, one must find similar 2D lattice with alternating D-A structure as well as the liquid of CT states. Compared to the conventional soliton lattice \cite{SolitonL1,SolitonL2}, there are two points of difference. One is the dimension and the other is the pairing of two fermions, which serve as the essential terms of our CT liquid theory.

We can now write down a toric-code-like Hamiltonian, that is,
\begin{eqnarray}\label{ZX}
\hat{H}=-\sum_p\epsilon_p\hat{A}_p-\sum_v\delta_v\hat{B}_v,
\end{eqnarray}
where $p$ denotes the plaquette with four D-A bonds (links), $v$ is for the vertex, $\epsilon_p$ and $\delta_v$ are the relevant energy, and
\begin{eqnarray}\label{ZX}
\hat{A}_p=\hat{Z}_1\hat{Z}_2\hat{Z}_3\hat{Z}_4,~~~\hat{B}_v=\hat{X}_1\hat{X}_2\hat{X}_3\hat{X}_4,
\end{eqnarray}
are the composite operators that include the product the corresponding $\hat{Z}$ and $\hat{X}$ operators of each link on the plaquette and vertex, as sketched in Fig.~\ref{Fig2}. As $\hat{A}_p$ and $\hat{B}_v$ have either zero or two overlap links, it is straightforward to prove that all $\hat{A}_p$ and $\hat{B}_v$ commute with each other. There then emerge extensive local conserved quantities, which is the most appealing point of the toric code model. In details, the eigenvalues of $\hat{A}_p$ are still twofold, namely $+1$ for even number of CT states on each plaquette and $-1$ for odd number. So is $\hat{B}_v$. The system thus possesses local $\mathbb{Z}_2$ symmetry, and there could exist numerous CT states in the ground state which is why we call it as ``CT liquid". Herein, if we take $\epsilon_p$ always to be positive, the liquid of even number CT states is recognized as ground state, and the odd number as the excited state. This excited state can be named as ``CT exciton".

Similar with the generic mechanism of anyons, the CT exciton here can be moved between neighbouring plaquette by flipping the shared link in between. Through continuously flipping, namely switching between neutral and CT states on the links, the CT exciton freely move on the lattice without inducing additional excited states. This implies the motion of CT excitons is nearly adiabatic. Actually, each flip of the link is nothing but that the electron on D unit transfers to the paired A unit. After continuously flipping electrons are accumulated on these A units which is what we called ``self-accumulated charges" \cite{Yao18}. It means the electron and hole in the CT state do not move such far; they just hop back and forth between two neighbouring units. Therefore, we can argue the thermodynamic entropy during this process does not essentially increase. The CT exciton merely deliver local energy and quantum information, which is why we call it ``exciton". We remark here that, this might serves as the actual channel for coherent energy transfer and thus the efficient charge photo-generation in nonfullerene solar cells.

\subsection{Electronic Wigner crystals and molecules}

Readers may be doubted that, the organic materials are normally regarded to be with strong disorders, so how can they form such an ordered lattice as in our toric code model? In order to unveil it, let us consider the so-called electronic crystals.

In condensed matter physics, the model of Wigner crystals expresses a scenario in systems with low-density electrons and strong two-body interactions \cite{Wigner1}. When the distance between two electrons is sufficiently long, the Coulomb interaction following $1/r$ scaling will dominate the kinetic energy with $1/r^2$ scaling, so the electrons will be localized and form an electronic crystal \cite{Wigner2}. A number of recent experiments have demonstrated the appearance of Wigner crystals in low dimensions \cite{Wigner3,Wigner4,Wigner5,Wigner6}. Furthermore, if the size of Wigner crystal is limited, it becomes the so-called Wigner molecule which is also observed in few-electron quantum dots and so on \cite{Wigner7,Wigner8}. One of the most remarkable effects in Wigner crystals is $4k_{\rm F}$ Wigner oscillations stemming from the pairing correlations of electrons. The quantum conductivity turns out to be the main signature of Wigner crystals, which can also be observed in soliton lattices \cite{SolitonL2}. Ferromagnetism can be found as well in terms of optimized doping \cite{Wigner9}.

Essentially, Wigner crystals and molecules are many-body localized systems without disorders. Two-body interactions are the most important ingredient, which easily leads us to think about the Coulomb attractions between electron and hole in CT states. Considering the kinetic energy in organic materials is normally small, our CT liquid model is thus closely equivalent to the Wigner molecules. It is worth noting that, in diluted magnetic semiconductors or magic-angle graphene, the background electrons are delocalized and the localized states are generated by external artificial treatments. On the contrary, electrons in organic materials are naturally localized by self-trapping polaron effect, so the emergency of Wigner molecules could be even easier than in other materials. As a consequence, although the organic molecules are commonly in the amorphous phase, the electronic states can separately form ordered lattice independent on the relative positions of atoms. This serves as an explanation that the CT liquid can be robust against the strong disorders on the molecular level.

\subsection{Phonons}

As stated, phonons are normaly the primary source of decoherence. Although the CT excitons seem to be undissipative, the situation becomes different while phonons are taken into account. We now have to unveil how the coherence of CT excitons survives the phonons and thus the warm and wet environment.

Phonons can be classified into two categories. One is the primary phonons on the dynamic level. They are non-perturbative and normally dealt with quantum dynamic approaches. The other is the secondary phonons on the thermodynamic (kinetic) level. They are perturbative and can be treated by molecular dynamic approaches. In traditional semiconductors the latter is more important. But organic semiconductors are mainly composed of carbon, hydrogen, oxygen and nitrogen. They are light and easy to couple with electrons to induce strong vibronic couplings, so the former one has also to be considered comprehensively. We notice that, in molecules with large weight, one might be thinking the spectral function of vibrational modes should be quite diverse, but in the actual cases merely several primary modes play essential roles by influencing the electron dynamics, especially in rigid molecules. The other modes can be recognized as secondary modes which solely act as thermal reservoir. In this context, our discussions can be separated into twofold. Within a timescale shorter than nanosecond, the secondary phonons almost do not work so we can merely take the primary ones into consideration. The mesoscopic quantum effects discussed below follow this line. On the other hand, if we want to go in depth into more macroscopic scales, i.e. longer than microsecond, it is necessary to consider a robust mechanism that the secondary phonons do not matter and the effect of primary ones is positive for quantum effects.

Phonons can also be classified into diagonal and off-diagonal modes, as stated. In our previous studies in both charge transport and CT processes \cite{Yao16,Yao17}, we found even if the off-diagonal couplings dominate the electron is not fully delocalized to all molecules but just dispersive in a finite extent, so that it is not precise to say the system is in a delocalized phase. We may call it as the ``dispersive phase" instead of the delocalized phase for a precise terminology. Actually, this dispersive characteristics has been observed by the electron spin resonance (ESR) experiment, which showed that the spatial extent of electrons in pentacene is roughly ten molecules \cite{10m}.

\begin{figure}
\includegraphics[scale=0.45]{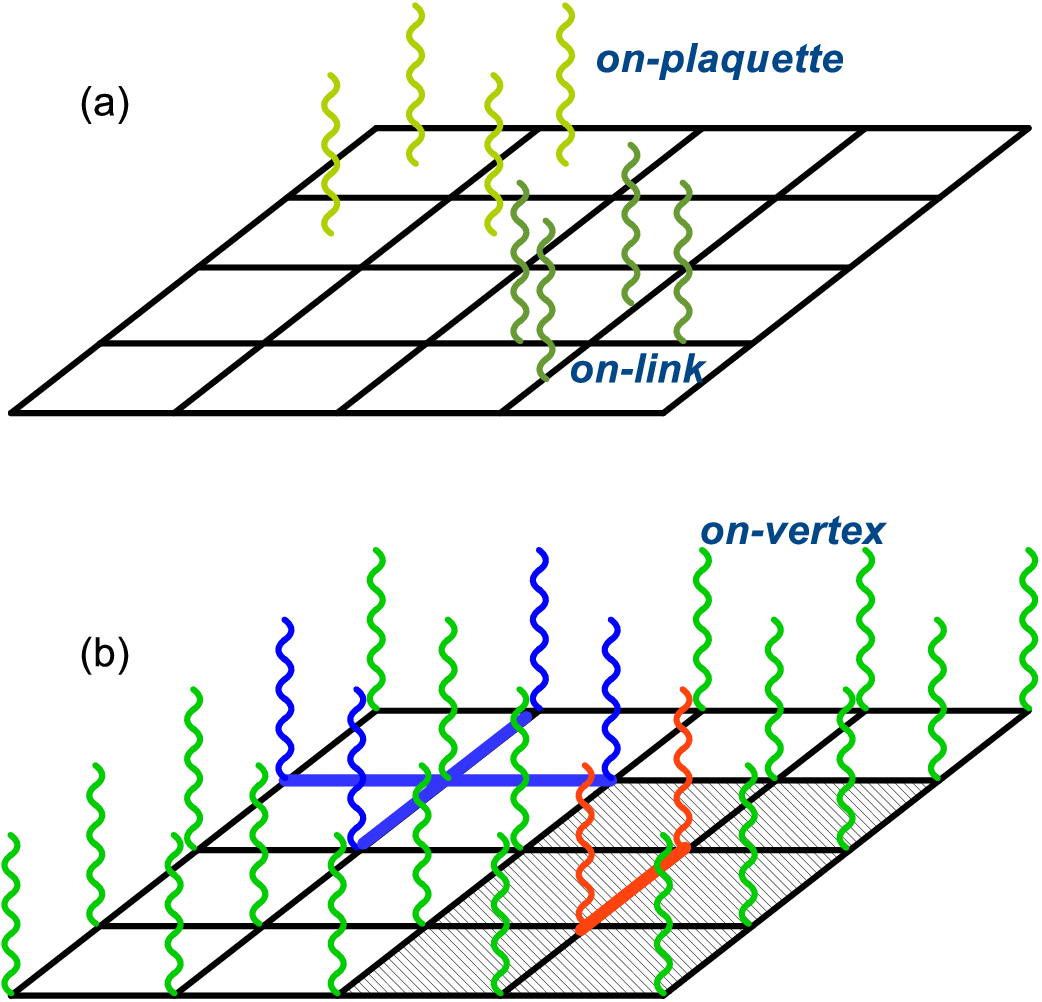}
\caption{Schematic for the CT fracton model. (a) Phonons denoted by wavy lines are on-plaquette for diagonal modes and on-link for off-diagonal modes. (b) On-vertex phonons are induced by aggregate which can couple to link operators (orange) or collectively to vertex operators (blue).}\label{Fig3}
\end{figure}

As shown in Fig.~\ref{Fig3}(a), the two phonon modes are respectively depicted by the wavy lines on-plaquette and on-link. Each D-A bond in our toric code model separately couples to two on-plaquette (diagonal) modes and one on-link (off-diagonal) mode. In terms of the fracton model for polarons \cite{Fracton0}, one can use a Hamiltonian to some order to approximately describe the free motion of CT excitons. In Section IV below, we will discuss it in more details. Here we'd like to emphasize that, quantum effects are quite generic in nature so the model can not be too specific. Hence, we will not try to write down an explicit form of Hamiltonian but just comprehend the role of phonons into more depth.

\subsection{Aggregate phonons}

\begin{figure}
\includegraphics[scale=0.45]{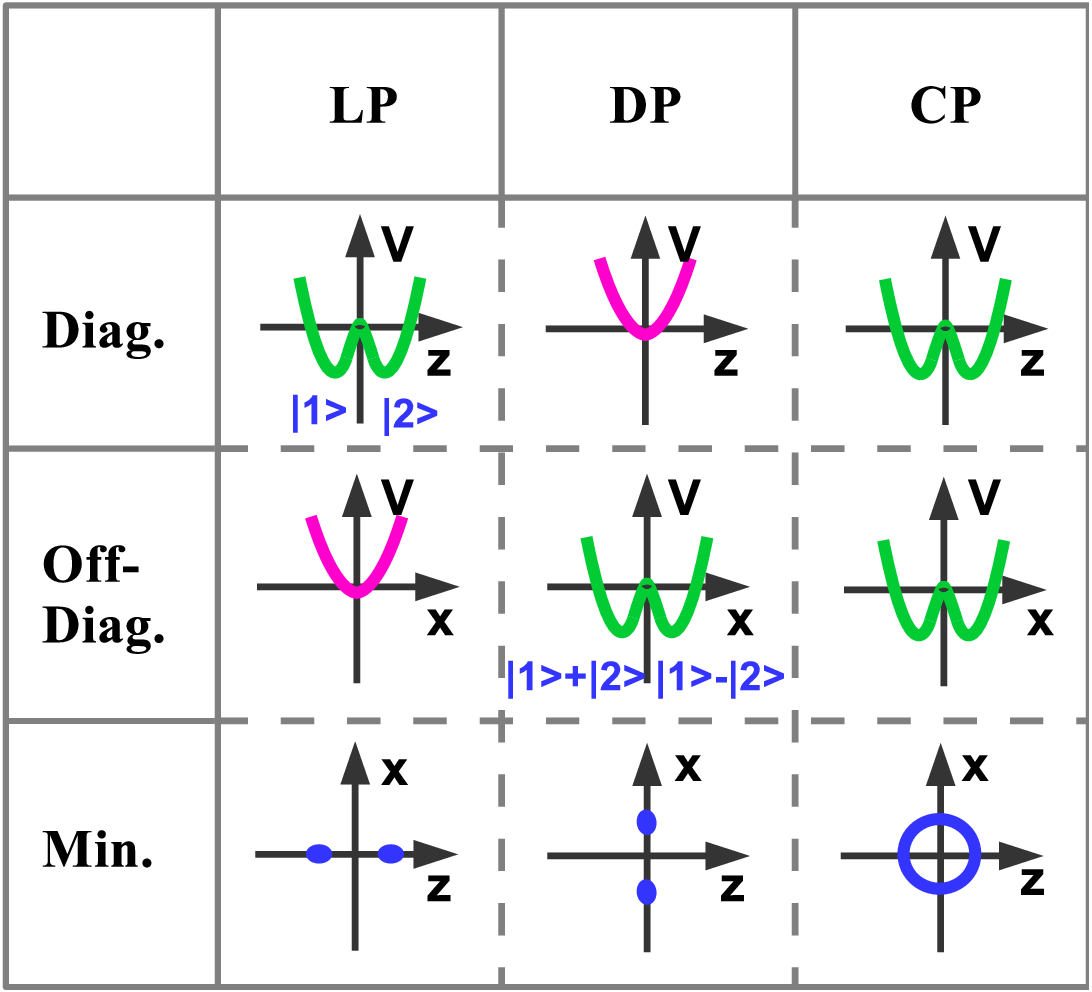}
\caption{Schematic for molecular potential surfaces in localized (LP), dispersive (DP) and critical (CP) phases. The generalized coordinate is set to $z$ direction for the diagonal (diag.) coupling and $x$ direction for the off-diagonal (off-diag.) coupling. $|1\rangle$ and $|2\rangle$ are the local states on units labeled as 1 and 2. The minimums (min.) of the potential surface are plotted on $z-x$ plane as well.}\label{Fig4}
\end{figure}

In a theoretical manner, the most simplest but highly nontrivial model to investigate the role of phonons is the so-called spin-boson model \cite{SBM1,SBM2}, which is extensively utilized to the researches of localization and dispersion of electrons. In the study of two-bath spin-boson model, a new phase named critical phase was uncovered \cite{Yao14}. When the diagonal and off-diagonal couplings are equal, it can be proven that the ground state of the system is highly degenerate due to the transition from parity symmetry to (nearly) continuous symmetry, as depicted in Fig.~\ref{Fig4} with a scheme of the molecular potential surfaces in three phases. We take D and A units labeled as 1 and 2 as instance and set the generalized coordinate on $z$ direction for diagonal couplings and $x$ direction for off-diagonal couplings. In the localized phase, the minimums of the potential surface appear on the $z$ axis and entangle with the local states on unit 1 and 2, respectively. In the dispersive phase, they appear on the $x$ axis and entangle with two dispersive states, namely the bonding state $|1\rangle+|2\rangle$ and the anti-bonding state $|1\rangle-|2\rangle$. On the lattice of more units, the dispersive states become superposition of Bloch waves. Of more interests is the critical phase with equal diagonal and off-diagonal couplings. An infinite number of minimums form a circle on the $z-x$ plane and the system can shift between localized and dispersive without any energy barriers, which may lead to exotic phenomena. In realistic materials, diagonal and off-diagonal couplings can not be explicitly the same. But so long as they are close to each other, there will be numerous local minimums in the high-dimensional potential surfaces resulting in the similar effects with that in the critical phase \cite{YaoSR}.

What makes the two couplings close is the resonant phonons in aggregate. It is well known that aggregate can induce many appealing effects such as aggregation-induced emission (AIE) \cite{AIE} which is absent in solution phase. In aggregate phase, many low-frequency vibrational modes are frozen so that channels of thermal dissipation are greatly suppressed. In our understanding, aggregate phonons become collective or resonant in many molecules. It is thus hard to figure out the aggregate phonons are diagonal or off-diagonal; they are in critical phase. Considering four units of two D's and two A's in a square, the diagonal operators $\hat{Z}$ couple to the four on-plaquette phonons while the off-diagonal operators $\hat{X}$ couple to the four on-link phonons. Intuitively, we can image wavy lines on the vertex to represent the composite aggregate phonons instead of on-plaquette and on-link, as sketched in Fig.~\ref{Fig3}(b). Each on-vertex phonon correlate with four nearest D and A units. We argue that, the more rigid the backbond of aggregate is, the more important the on-vertex phonons are.

Since the aggregate phonons stem from the collective competition among various vibrational modes, it is hard to write down an explicit Hamiltonian for the vibronic couplings between CT excitons and these phonons. In Section IV, we will consider three realistic models involving competing interactions to study whether the aggregate phonons induced high degeneracy and critical phase give rise to robust quantum effects. Here, we solely describe a schematic picture for these phonons.

As stated, once the system is in the critical phase, no matter the spin is along which direction, the phonons are always polarized due to the high degeneracy. This is just to say the aggregate phonons always hold a non-zero displacement $D$ or $-D$. Let us denote the twofold ground states of phonons as $|D\rangle$ and $|-D\rangle$ and suppose all the aggregate phonons on the lattice residing in $|D\rangle$ is the ground state of the whole system, without loss of generality. An $\hat{A}_p$ operator corresponds to four on-vertex phonons, and a $\hat{B}_v$ operatore refers to five, as shown in Fig.~\ref{Fig3}(b). If, in a generic sense, the operation of a single $\hat{Z}$ or $\hat{X}$ operator is followed by flipping from $|D\rangle$ to $|-D\rangle$ of two correlated on-vertex phonons, the action of $\hat{A}_p$ does not change the parity of phonons but the action of $\hat{B}_v$ flips the sign of displacements (parity sectors) of phonons on the four terminals of the vertex. In order to keep the system still in the ground state, it is necessary to assume that even number of phonons residing in $|-D\rangle$ state does not change the energy. Again, we have got two eigen-states of the whole system: the ground state in which total number of CT states and $|-D\rangle$ states of aggregate phonons is even, and the excited state in which the total number is odd. This perspective is strongly motivated by the X-cube model of fracton \cite{Fracton1}, so we can call the excited states on this lattice as ``CT fractons".

Similar with that in the fracton physics, the motion of CT fracton is also dimension limited. As sketched, a single $\hat{X}$ term couple with two aggregate phonons will induce six fractons, suggesting the motion of CT excitons can not be adiabatic in this situation. It is however interesting that, if we just flip parity sector of a single aggregate phonon, two neighboring CT fractons are generated. Continuously flipping the other phonon's sector, two bound CT fractons can then be freely moved together along the normal direction. This effect is very similar with the dimension-limited mobility of fractons \cite{Fracton2}. Only the first change of phonon's parity sector has to be activated by thermal fluctuations and the following flips do not cost thermal energy, so this is a thermally triggered coherent motion of CT fractons. In realistic materials, there are many quantum effects relating to the thermal energy, such as the thermally activated delayed fluorescence (TADF) \cite{TADF} and endothermic singlet fission \cite{Endo}. As stated, no matter diagonal or off-diagonal, phonons and thermal fluctuations are intuitively regarded to be negative for quantum effects. Here, borrowing the picture of fracton physics, we propose a generic mechanism that the role of aggregate phonons can be positive to the undissipative motion as well as the robust quantum effects. In the following Section, we will go deeper into realistic materials and models to examine if the mechanism is actually available.

\section{Materials}

As mentioned, single crystal phase is rarely present in organic semiconductors, and the amorphous film serves as the primary form. The quality of the aggregate structure thus completely depends on the self assemble and organization. The intrinsic interactions among molecules emerge as the sole ingredient. Physical approaches such as doping, injection and compression can not help significantly enhance the performance of materials. In order to enable or enhance some functions, one has to search for appropriate organic intermediates and rationally modify them on both side and end groups. In this context, we can divide the quantum effects in organic semiconductors into three levels: molecule, aggregate and bulk. This turns out to be a hierarchy of quantum effects \cite{Wigner8}, as listed in Table \ref{Tab}.

\subsection{Hierarchical quantum effects}

\begin{table*}
\centering
\caption{The hierarchy of quantum effects in organic semiconductors}
\label{Tab}
\begin{tabular}{c|ccccc}
  \toprule
   & \textbf{Morphology} & \textbf{Number of molecules} & \textbf{Phenomenology} & \textbf{Mechanism} & \textbf{Material candidate} \\
  \colrule
   & & & Bandlike mobility & & Crystalline acenes \\
  \textbf{Microscopic} & Molecule & Fewer than thousand & Magnetoresistance & Single carriers & Alq$_3$, CuPc, etc. \\
   & & & $\dots$ & & $\dots$ \\
  \hline
    & Aggregate & & Coherent hole transfer & & Nonfullerene acceptors \\
  \textbf{Mesoscopic} & Nanorod & Millions & Singlet fission & CT excitons & Imide-fused hydrocarbons \\
   & Cluster & & $\dots$ & & $\dots$ \\
  \hline
   & & & Superconducting & & K$_3$C$_{60}$ \\
  \textbf{Macroscopic} & Bulk & Infinite & Ferromagnetism & CT fractons & C$_{60}$(TDAE)$_{0.86}$$^{\rm a}$  \\
   & & & $\dots$ & & $\dots$ \\

  \botrule
\end{tabular}\\
\footnotesize{{\rm a} CT salt of fullerene and tetrakis(dimethy1amino)ethylene}\cite{fuller2}\\
\end{table*}

If the molecules have got a strong intermolecular repulsion stemming from, e.g., the steric hindrance, the micro-structure tends to be disordered and the conjugation length for delocalized $\pi$ orbitals is quite short. For example, the spatial extent of the wavefunction in pentacene has been determined to be ten molecules by ESR measurement \cite{10m}. Although acenes are normally called as organic crystalline material, the great degree of disorder makes the electrons localized within limited scale. Similar results can be found in fullerene and other polymers \cite{Delo5,Delo8}. It is worth noting that, although the molecular weight of polymers could be very large, the conjugation length is still short due to the ubiquitous entanglement and twirling of the chains. Actually, as widely accepted, the disorders are always present and even dominant, so it is obviously difficult for the quantum effect to survive. To find a mechanism against the overwhelming disorder and decoherence was indeed the initial motivation of this work.

As a consequence, even if there is some kind of quantum effect in these disordered systems, it should be relatively weak and within short region. We denote it to be \textit{microscopic} quantum effect. The typical phenomena refer to bandlike charge transport in crystalline materials and magnetic field effects, such as magnetoresistance and magnetoelectroluminescence in small molecules like Alq$_3$ (q for quinine). As our previous works indicated \cite{Yao12,Siwei}, both these two phenomena stem from the following mechanism. When an electron goes to a molecule, it will stand there and wait for the expansion of its wavefunction. The quantum effect grows up until the wavefunction is dephased at the decoherence time $t_{\rm d}$. Another round begins and the entropy is continuously accumulated. In terms of the models in the last Section, one can reasonably expect this non-adiabatic quantum effect of single carriers must be within short scales even individual molecules. On the experimental side, it is difficult to directly detect the quantum coherence in spatial dimension, and one can merely know it from indirect measurements such as $\mu-T$ relationship.

In order to obtain observable quantum effect, one has to synthesize materials with very nice aggregate structure. This structure must benefit from the characteristics of molecules themselves. For example, the imide-fused hydrocarbons such as perylene diimide (PDI) molecules have got perfect planar structure and will naturally form nanorods through self-assemble in special solvent with the size being several micrometers. The intermolecular (two-body electron-hole) attraction is so large that even in the bulk heterojunction PDI would still form a cluster (Wigner molecule) making the film phase separation. The nanorod is extremely rigid and does almost not dissolve in common solvents such as water and alcohol. As discussed above, the rigidness of the nanorods stem from A-D-A structure of PDI molecule and the induced dipole interactions. In terms of our model here, this structure easily induces CT excitons which can adiabatically deliver local energy in principle. Meanwhile, the rigidness may also enable multi-molecular resonant vibrational modes further enhancing the protection of quantumness and counteraction of entropy increase. This rigid and ordered structure in a mesoscopic scale also appear in nonfullerene solar cells. It gives rise to \textit{mesoscopic} quantum effects such as coherent energy and hole transfer which are nowadays recognized as the main mechanism for high efficiency of charge photogeneration \cite{Yao19}.

By definition, the \textit{macroscopic} quantum effect in condensed matter physics emerges in the scales that human beings are living. Both the superconducting and ferromagnetism have been observed in organic systems such as fullerene doped by potassium and TDAE (tetrakis(dimethy1amino)ethylene) \cite{fuller1,fuller2}, but the critical temperature is still low. Although the mechanisms are so far not clear, CT states apparently play essential roles in these systems, especially in CT salts. If people want to further increase the critical temperature, a rational strategy of design in our opinion is bottom-up, namely to continuously enlarge the mesoscopic quantum effects. A top-down strategy seems not work because the material candidates are limited. Our recent work has revealed the appearance of ferromagnetism at room temperature PDI anions \cite{FOS}, and we strongly expect more macroscopic quantum effects such as superconducting can be observed in the similar systems.

\subsection{Polycyclic aromatic hydrocarbons}

Polycyclic aromatic hydrocarbons (PAHs) long-termly serve as the main intermediates in organic synthesis due to the good conjugation of $\pi$ orbitals and well-understood aggregate structures. In photophysics, the H- and J-aggregation give rise to completely different transition dipole moments, so the charge photo-generation in these two structures are different. This commonly accepted perspective bases upon classical electric dipoles and quantum effects are not considered. Obviously, PAHs act far more than just as a classical system.

According to Clar's aromatic $\pi$-sextet rule, fewer number of sextet rings gives rise to higher activity, and people are usually interesting in those PAHs with chemically instable molecular structures \cite{Radical}. $\pi$ electrons in them are easy to form open-shell radicals or generate $\pi-\pi$ bonds with other molecules. Radicals exist in almost all organic materials, and are greatly grown in quantity by redox reaction. In absence of kinetic blocking, radicals are nearly free to (chemically) delocalize in a certain extent to lower the extremely high chemical activity. Readers may ask a question: Is there any difference between radicals and polarons? This has to be carefully justified but is not the main scope of the present work, and a recent relevant work is recommended for the interesting readers \cite{recent}. Consequently, radicals are the essential mediate for the generation of CT liquid and thus significant in many quantum effects such as ferromagnetism.

$\pi-\pi$ stacking turns out to be the essential structure of PAHs for the emergence of exotic phenomena. The length of the $\pi-\pi$ bond is around 3-4 {\AA} and the bonding energy is slightly weaker than that of covalent $\sigma$ bond. Different from the covalent bonds in polymers, the $\pi-\pi$ stacking is normally confined by, e.g. steric hindrance, so the structure is more rigid than that of polymers. Via self-assemble, parallel stacking structure in perylene and herringbone aggregate in pentacene can be organized. The structures also determine the emergence of quantum effects in PAHs.

Since radicals naturally exist, chemists have to do something to lower the instability during the synthesis. The most common approach is to modify the end groups, such as fusion by imide. The nitrogen on the imide prefers to form a hydrogen bond with the carbonyl group on the other imide, very similar with that in DNA. The hydrogen bond can further help stabilize the rigid structure, but can also be dissociated by specific solvent. As a result, one can rationally add more functional groups into the end groups to get more appealing effects. The backbone of PAHs and the end groups then form a standard D-A heterojunction in the rigid $\pi-\pi$ stacking structure. Our CT liquid model is then substantiated in these materials.

If the number of sextet rings is sufficiently large, we can also call these PAHs as graphene nanoplatelets. The researches of macroscopic quantum effects in graphene nanoplatelets also emerge to be inspiring, but this is out of our current scope.

\subsection{Metal-organic complexes and charge-transfer salts}

While talking about metal-organic complexes, one may at the first glance think about metal-organic frameworks, but here we mainly focus on the organic small molecules. In the practical applications, metal-organic complexes such as Alq$_3$ and CuPc (Pc for phthalocyanine) manifest their pioneering value especially in light-emitting devices. There are numerous advantages for them in synthesis, fabrication and encapsulation. On the current stage, one would ask if it is still valuable to further investigate them given the existence of established industry. This is determined by whether there are novel quantum effects.

As stated, metal-organic complexes normally manifest weak magnetism and then magnetic field effects emerge in these materials, e.g. Alq$_3$. However, since these effects are on the microscopic level, they are difficult to be utilized in practical applications. People then tried to apply the molecular qubits on quantum computations. A good example is the dyad of CuPc and N@C$_{60}$ \cite{MOC1}, in which CuPc behaves as the D unit and fullerene as the A unit as usual. The ESR spectroscopy is able to separately feature the spin configurations of fullerene and copper in CuPc. Via chemical modifications on the spacer groups between these two units, it is possible to realize the quantum control of spin qubits in the molecular systems. Similar effects can also be observed in organic CT salts. For example, in the D-A-radical system \cite{MOC2}, researchers realize the quantum teleportation among these different units driven by photons. All these experiments convincingly exhibit the promising potentials that these D-A hetero-structures can be utilized in quantum applications.

The role of metals is not just the donor of electron and spin. Recent work has revealed the essential function of catalysis \cite{Cata}. Actually, if one wants to obtain a good D-A structure, hydrogen plays significant role. Our quantum chemistry computations show the adsorption of hydrogen are the main source of radicals and thus the CT excitons, which will be published soon. Hydrogen would not initiatively escape the original dopants such as (4-(1,3-dimethyl-2,3-dihydro-1H-benzoimidazol-2-yl)phenyl)dimethylamine (N-DMBI-H), hydrazine hydrate and so on, so the catalysis of metals turns out to be irreducible. Are there any quantum effects in this hydride transfer process still needs more comprehensive investigations.

Metal-organic complexes such as K$_3$C$_{60}$ and CT salts such as C$_{60}$(TDAE)$_{0.86}$ have attracted research interests in superconducting and ferromagnetism fields for many years  \cite{fuller1,fuller2}. However, the research progress of macroscopic quantum effects in organic materials goes much slower than other fields. A possible reason is the mismatch of physical and chemical treatments. Physicists prefer to explore better crystalline structure while chemists devote effort to modify the molecular structures. As described in our CT fracton model, both the molecule and aggregate structures are equivalently important. Balance between these two perspectives will gain great successes.

\section{Dynamics}

The toy model describes a phenomenological scenario of CT fractons to unveil the robustness of quantum effects. One would be doubting that in the realistic devices the situation can not be as ideal as expected in theory. In our opinion, given the D-A alternating structure which is commonly present in the nowadays synthetic materials as introduced in the last Section, the key issue is the multi-molecular resonant vibrational modes, i.e. aggregate phonons in our terminology. As discussed, when different phonon modes compete with each other, the critical phase emerges and the role of phonons probably becomes positive. In this context, we have to carefully investigate the dynamics of CT states in phonon baths on the basis of more practical models other than toy models.

On the experimental side, the state-of-the-art technique for detecting CT states is to measure the luminescence, which especially needs a bright CT excited state. In our present theory, CT states play more significant roles than just as an excited state. Therefore, it is necessary to propose a more generic scheme for detecting CT states, no matter excited or ground state. We notice that OTOC is a nice quantity that can be straightforwardly applied to the CT states, in the sense of both measurements in practice and understanding in mechanisms. Therefore, we conduct a series of computations of OTOC for different dynamic systems in organic semiconductors.

\subsection{OTOC}

Recently, following the rapid developing progress of theories \cite{MBL0,MBL1,MBL2,MBL3,MBL4,MBL5} and experiments \cite{OTOCexp1,OTOCexp2,MBLNew1,MBLNew2}, a quantum dynamic quantity for measuring the entanglement entropy originally from the cosmology which is the so-called OTOC defined as \cite{OTOC1,OTOC2},
\begin{eqnarray}
F(t)=\langle \hat{W}^{\dag}(t)\hat{V}^{\dag}(0)\hat{W}(t)\hat{V}(0)\rangle.
\end{eqnarray}
Herein, the expectation value is averaging over the eigen-state of the Hamiltonian and $\hat{W}(t)=\exp({i\hat{H}t/\hbar})\hat{W}\exp({-i\hat{H}t/\hbar})$. $\hat{W}$ and $\hat{V}$ are two local operators which commute with each other at time 0, and $\hat{W}^2=\hat{V}^2=1$. For the CT states in our theory, e.g., $\hat{W}$ and $\hat{V}$ represent nothing but the $\hat{X}$ and/or $\hat{Z}$ operators at different units. Physically, the OTOC quantifies both the spatial and temporal correlation of two initially local operators \cite{OTOC3,OTOC4,OTOC5,OTOC6}.

\begin{figure}
\includegraphics[scale=0.45]{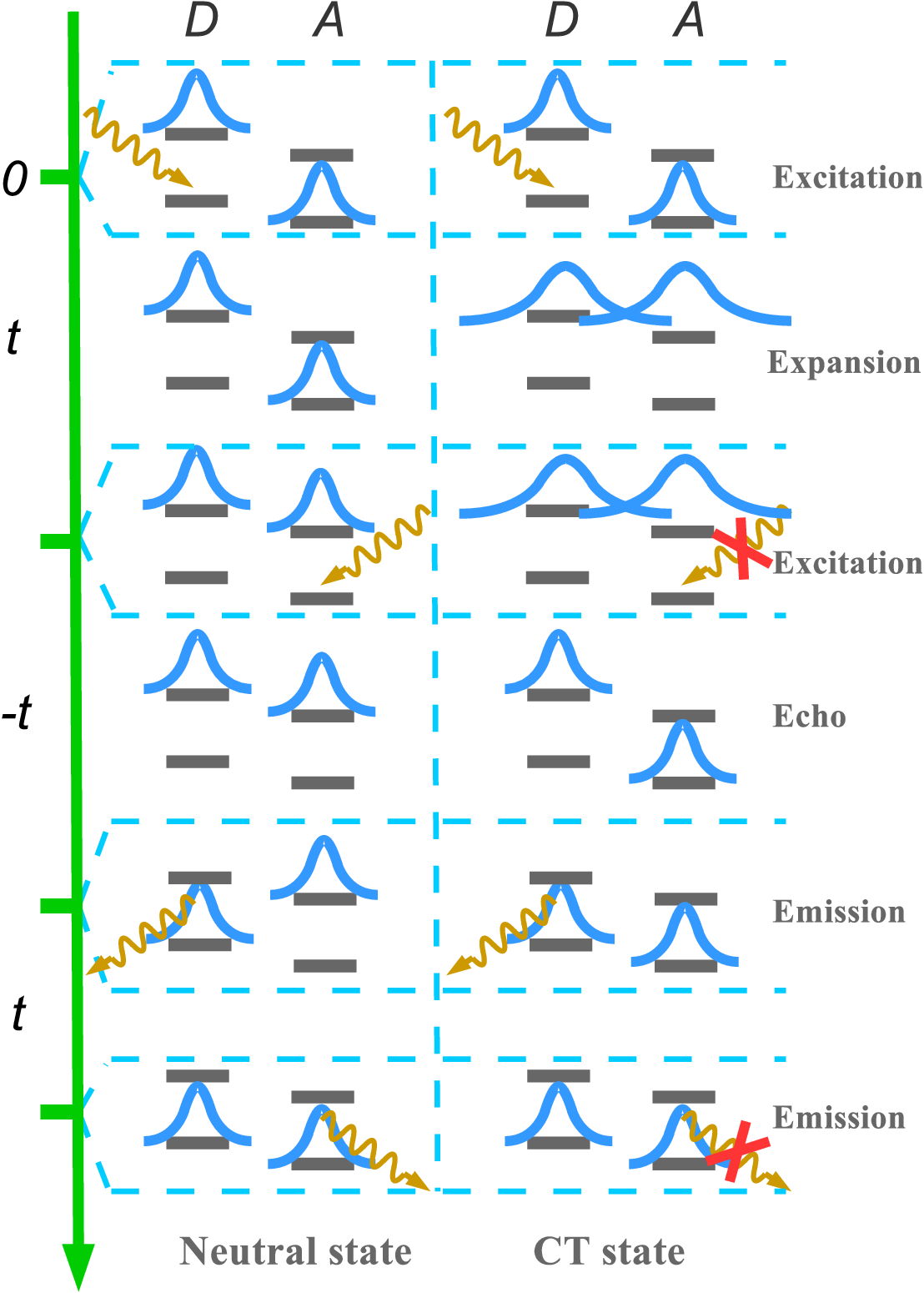}
\caption{Schematic for the physical scenario of OTOC for neutral and CT state. At time 0, the D unit is excited by a pulse. During the following time evolution, the wavepacket of exciton will be kept unchanged in neutral state while expanded to A unit for CT state. At time $t$, the A unit is excited in neutral state while it is not able to absorb the light as the exciton has been transferred to it before the laser pulse comes. Afterward, the system evolves to time $-t$ via an echo dynamics. In experiments, this can be realized through reversing the sign of all parameters in the Hamiltonian. In the echo duration, the neutral state do not change, but the expanded wavepacket is shrunk to its original form in the CT state. Subsequently, in the following emission events at time 0 and $t$, one obtains different emission signals for the two states.}\label{Fig5}
\end{figure}

To let the nonspecialists easily understand and make sense, Fig.~\ref{Fig5} sketches a scenario that how the OTOC can be applied to detect CT states. At time 0 we excite the D unit by a pulse of laser, microwave and so on, and then let the system evolve. After time $t$ we excite A unit and let the system evolve back to time 0 (echo dynamics). Then we measure the emission signal of D and let the system again evolve to time $t$ and finally we measure the emission signal of A. If the system is fully localized, the excitations on D and A are independent with each other so that the OTOC is definitely 1, namely the emission signals of these two units are always completely detectable and do not form a CT state. On the other hand, if the system is fully delocalized, the exciton on D can be quickly transferred to A to form a CT state before time $t$, so that the A can not be repeatedly excited and the OTOC in this case decays exponentially from 1 to 0. The critical phase is in between. Due to the finite dispersive extent, there is a certain probability that the CT state is generated, so the OTOC normally oscillates between 0 to 1 showing an uncertainty feature \cite{OTOC3,OTOC4}. It has been demonstrated that the OTOC is closely related to the second R\'{e}nyi entropy \cite{OTOC5}.

There are many theoretical measures for the entanglement entropy. One may ask why we do not use the simpler quantities such as the von Neumann local entropy and the inverse participation ratio (IPR), as the entropy is always mentioned in the charge separation processes \cite{Entropy1}. The advantage of using OTOC as an indicator is that, the OTOC as an explicit function of real time is specifically developed for characterizing the \textit{quantum dynamics} of spatial localization and dispersive feature in closed systems. Especially, in the dispersive process the vibrational modes are essential in a quantum dynamic manner while in localized phase they are just a medium for heat exchange, and only the OTOC is able to make a clear distinction between them. So the calculation of OTOC may give rise to an insightful scenario to understand the role of phonons, especially those in aggregate \cite{osci1,osci2,osci3,osci4}. Furthermore, compared with the local entropy which is defined in the equilibration, OTOC is a purely dynamic quantity that can be easily measured by the commonly-used ultrafast spectroscopy \cite{OTOCexp2}. On the other hand, in presence of translational symmetry, the IPR can be irrationally large in the localized phase as discussed below, making it difficult to distinguish it from the critical phase. Whereas, they are quite easy to be distinguished by OTOC which is symmetry insensitive. Consequently, it is remarkable to calculate the OTOC as the measure of entanglement entropy in organic semiconductors to examine the actual roles of CT excitons and aggregate phonons. It is also worth mentioning that, the quantitative connection between the entanglement entropy and the thermal entropy in the statistical mechanism is on the current theoretical stage not straightforward except in some specific cases, so the more explicit quantification for the real devices is beyond our present scope.

\subsection{Ultrafast CT process}

In the normal scenario of organic photovoltaic devices, upon photoexcitation a local exciton is generated followed by a charge separation process to produce photocurrent. The charge separation comprises two steps: The first is the charge transfer between donor and acceptor and the second is the further dissociation of the CT state. Borrowing the language of traditional semiconductors, the CT state is usually regarded as a tight-binding electron-hole pair with the binding energy mainly from the Coulomb attraction. Considering the relatively small dielectric constant, however, a puzzle arises: what serves as the driving force to make the CT state dissociate. As stated, recent experiments \cite{Cohe1,Cohe2,Delo1,Delo2,Delo3,Delo4} highlight the quantum coherence and the emergence of the long-range CT states which possess much smaller binding energy than the short-range ones. In this context, it is thus unjustified to define the CT state by the Coulomb interaction. Our CT exciton model provides another perspective.

On the other side, the photophysics of organic photovoltaic devices is often discussed with a quantum heat engine model. In thermodynamics, an ideal cycle comprises four processes: isothermal expansion (heat injection $Q_1$), adiabatic expansion (power export), isothermal compression (heat rejection $Q_2$), adiabatic compression (internal energy recover). The efficiency is defined as $\eta=1-Q_2/Q_1$. In the generation of local exciton, a single excited state (S$_1$ state) is populated and the entropy increases. This is obviously the heat injection process, and $Q_1$ could be simply regarded as the energy of the absorbed photons. As discussed in the toy model, the motion of CT excitons triggered by aggregate phonons is approximately isentropic and adiabatic processes which could be regarded as the process of power export. During the process that the local exciton is conversed to the CT state the entanglement entropy increases which will be quenched in the following CT-state dissociation and charge separation process. Combined with the energy relaxation from higher-energy excitation to the lower one, these two processes serve as the major channel of heat rejection and determine the value of $Q_2$. The subsequent rebalance of the electrostatic potential with the external circuit can be regarded as the final adiabatic process of the cycle of quantum heat engine.

To rationalize the efficiency of the heat engine, it is essential to quantify $Q_2$, so the entropy change must be well measured. The transition from the local exciton to the CT state turns out to be essential. There are two ways to estimate the entropy. One is following the classical statistical physics, suggested by B. A. Gregg \cite{Entropy1}. Let us now assume the coherent size of CT state is $N$ molecules, and the electron and hole are dispersive in this region. Both of them are completely entangled to form an coherent state. As the local exciton is singlet, the (coherent) CT state must be singlet as well. If we further assume that once thermalized, the singlet exciton, the electron and the hole are respectively distributed on $N$ molecules with equal probabilities, the entropy of a singlet exciton $S_{\rm SE}$ should equal to $k_{\rm B}\ln N$ and the entropy of a CT state $S_{\rm CT}$ should be twice of it. The entanglement entropy that is quenched during the CT-state dissociation should thus be close to $k_{\rm B}\ln N$. Imagine a coherent sphere for the excitons with the radius being 10 molecules \cite{10m}, so $N$ should roughly equal to 4200. At room temperature, we thus obtain $Q_2=T\Delta S\simeq 217{\rm meV}$, which could be regarded as the heat rejection in the photo-electric conversion process. By this way of estimation of entropy, how to deal with the decoherence and thermalization is the central problem. The other way to estimate the entropy is to directly compute the entanglement entropy between electron and hole in a dynamic manner. In our previous work, we have calculated the ultrafast long-range charge transfer process taking the intermolecular vibrations into account. Therein, the entanglement entropy in a three-dimensional case is estimated to be 210meV \cite{Yao16}. This way of estimation allows us to deal with either coherent or incoherent dynamics in a unified framework. In addition, we believe the ultrahigh open-circuit voltage loss in organic photovoltaic devices may stem from this effect \cite{Voc1,Voc2}.

The above arguments rely on two prerequisites. Firstly, the excitons, electrons and holes must be quantum matters. There is no way the classical particles prefer to be confined in a relatively small region rather than to diffuse to much larger extent and produce unrealistically large entropy increase. The dispersive feature of the wavepackets however induces the spatial expansion of the coherent size. In order to confine the spatial size the system must be within a MBL phase, or in our language a critical phase. Secondly, the dispersive electrons and holes are driven by many-body interactions, such as the off-diagonal vibronic couplings \cite{Yao16} or electronic interactions \cite{Yao18}. These interactions ensure the entanglement entropy increases in a slow manner, such that the entropy does not significantly increase during the subsequent charge extraction process after charge separation. This picture is available in both fullerene and non-fullerene solar cells.

As discussed in our toy model, the emergency of critical phase gives rise to the formation of CT fractons. To examine whether this fracton scenario works in realistic materials, we first investigate the commonly-accepted Frenkel-CT mixed model, by which we have demonstrated the ultrafast long-range charge transfer process occurs \cite{Yao16}. We realized that the Coulomb interaction between electron and hole induces localization and the nonlocal vibronic couplings give rise to nonlocal disorder and dispersion, and they compete with each other to let the quantum effect emerges. The Hamiltonian writes
\begin{equation}
H=H_{\rm e}+H_{\rm p}+H_{\rm ep}.\label{hami}
\end{equation}
The first term $H_{\rm e}$ is written as,
\begin{eqnarray}
H_{\rm e}&=&E_{\rm ex}|0\rangle\langle0|+V_{\rm DA}(|0\rangle\langle -1,1|+{\rm h.c.})\nonumber\\&+&\sum_{i}\sum_{\{j,j'\}}V_{\rm A}|i,j\rangle\langle i,j'|
-\sum_{j}\sum_{\{i,i'\}}V_{\rm D}|i,j\rangle\langle i',j|\nonumber\\&-&\sum_{i,j}\frac{C}{r_{i,j}}|i,j\rangle\langle i,j|,
\end{eqnarray}
where $|0\rangle$ represents the state of a local Frenkel exciton and $|i,j\rangle$ is the CT exciton with the electron on A unit $i$ and hole on D unit $j$, $E_{\rm ex}$ is the on-site energy of the Frenkel exciton, $V_{\rm D/A/DA}$ denotes the electronic interactions, $C$ is the prefactor of the Coulomb attraction between electron and hole with $r_{i,j}$ being the distance which will induce the localization of electron and hole. We set the parameters as $E_{\rm ex}$=0.25eV, $V_{\rm DA}$=0.08eV, $C$=0.48eV, and $V_{\rm D}=V_{\rm A}$=0.06eV \cite{Yao16}. The phonon Hamiltonian $H_{\rm p}$ is expressed as ($\hbar=1$)
\begin{equation}
H_{\rm p}=\sum_{\nu}\omega_{\pm,\nu}\hat{b}^{\dag}_{\pm,\nu}\hat{b}_{\pm,\nu},\label{hamiph}
\end{equation}
where $\hat{b}^{\dag}_{\pm,\nu}$ ($\hat{b}_{\pm,\nu}$) denotes the creation (annihilation) operator of $\nu$-th phonon mode on the donor ($-$) or acceptor ($+$) molecules, and $\omega_{\pm,\nu}$ is the relevant frequency. For the vibronic couplings, we consider the off-diagonal couplings:
\begin{eqnarray}
&&H_{\rm ep}=\sum_{i,j,\nu}\gamma_{-,\nu}\left[|i,j\rangle\langle i-1,j|\hat{b}^{\dag}_{-,\nu}+|i-1,j\rangle\langle i,j|\hat{b}_{-,\nu}\right]\nonumber\\&&+\sum_{i,j,\nu}\gamma_{+,\nu}\left[|i,j\rangle\langle i,j+1|\hat{b}^{\dag}_{+,\nu}+|i,j+1\rangle\langle i,j|\hat{b}_{+,\nu}\right],\label{hamiexph}
\end{eqnarray}
with $\gamma_{\pm,\nu}$ being the coupling strength for $\nu$-th mode, depending on the spectral density $J(\omega)=\pi\sum_{\nu}\gamma_{\pm,\nu}^2\delta(\omega-\omega_{\pm,\nu})$. A continuous spectral density is set for both couplings which are chosen to be sub-Ohmic as usual, that is $J(\omega)=2\pi\alpha(\beta)\omega^{1-s}_c\omega^{s}{\rm e}^{-\omega/\omega_c}$ with $\alpha (\beta)$ being the dimensionless diagonal (off-diagonal) coupling, $s$ being the exponent, and $\omega_c$ being the cut-off frequency.

\begin{figure}
\includegraphics[scale=1.1]{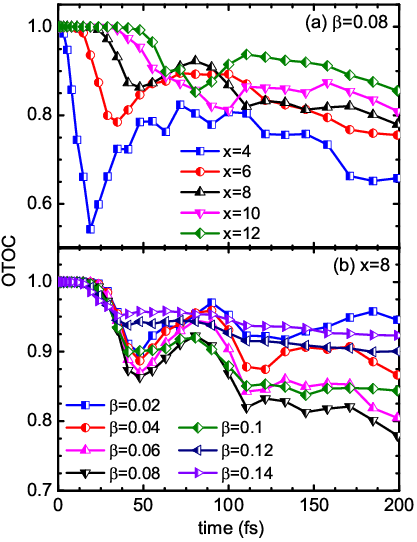}
\caption{Time evolution of OTOC in the Frenkel-CT mixed model as a function of (a) $x$ for $\beta=0.08$ and (b) $\beta$ for $x=8$.}\label{Fig6}
\end{figure}

In the previous work \cite{Yao16}, we have calculated the ultrafast dynamics of the long-range CT state, showing that for an intermediate vibronic coupling ($\beta=0.06$--$0.1$) the charge transfer is the most efficient. Moreover, we found that during the whole charge separation process, both the electron and hole modestly hold the shape of spatial confined wavepackets rather than further delocalize, suggesting that the ultrafast long-range CT process is similar with the case in critical phase. To examine this idea, we calculate the OTOC dynamics with the site index for the operator $\hat{V}(\equiv\exp(i\pi \hat{n}_1))$ being the first A unit. Fig.~\ref{Fig6} shows the results for various $x$ which is the unit index for the operator $\hat{W}(\equiv\exp(i\pi \hat{n}_x))$ and the off-diagonal vibronic coupling $\beta$. It is found that, the minimum value of OTOC transmits from the unit $x$=4 to $x=12$ during time evolving, exhibiting that the charges are transported along these units. Except the case $x=4$, the values of OTOC for other four cases are similar implying the long-range feature of the CT state. For the dependency of $\beta$, we can find for 0.02 and 0.04, the OTOC deviates from 1 slightly implying the localization dominates. From $\beta=0.06$ to 0.1 the OTOC oscillates between 0.7 and 1 rather than quickly decays to 0 indicating the Coulomb interaction functions. We do not calculate the longer time duration because it is rather time-consuming and the numerical precision is lost in the long-time evolution, but we expect the OTOC will finally decay to 0 in a very slow manner \cite{OTOC5}. The minimum appears at $\beta=0.08$, which is the case that the charge transfer process is efficient. Combined with the dynamics of population as shown in Ref.~\cite{Yao16}, $\beta=0.06$ to 0.1 is the regime that the long-range CT process takes place. In addition, for $\beta=0.12$ and 0.14, the case that charge transfer does not occur, the OTOC keeps nearly constant above 0.9. The coupling is so strong that the phonon energy exceeds the electronic band, so the electron is again localized. We can thus conclude here that the ultrafast long-range CT process in our model does work in the critical phase, and neither weak nor strong vibronic couplings are friendly to the quantum effects wherein --- the intermediate ones are best, which are fairly appropriate for the emergency of aggregate phonons and thus the CT fractons.

\subsection{Charge transport}

\begin{figure}
\includegraphics[scale=0.45]{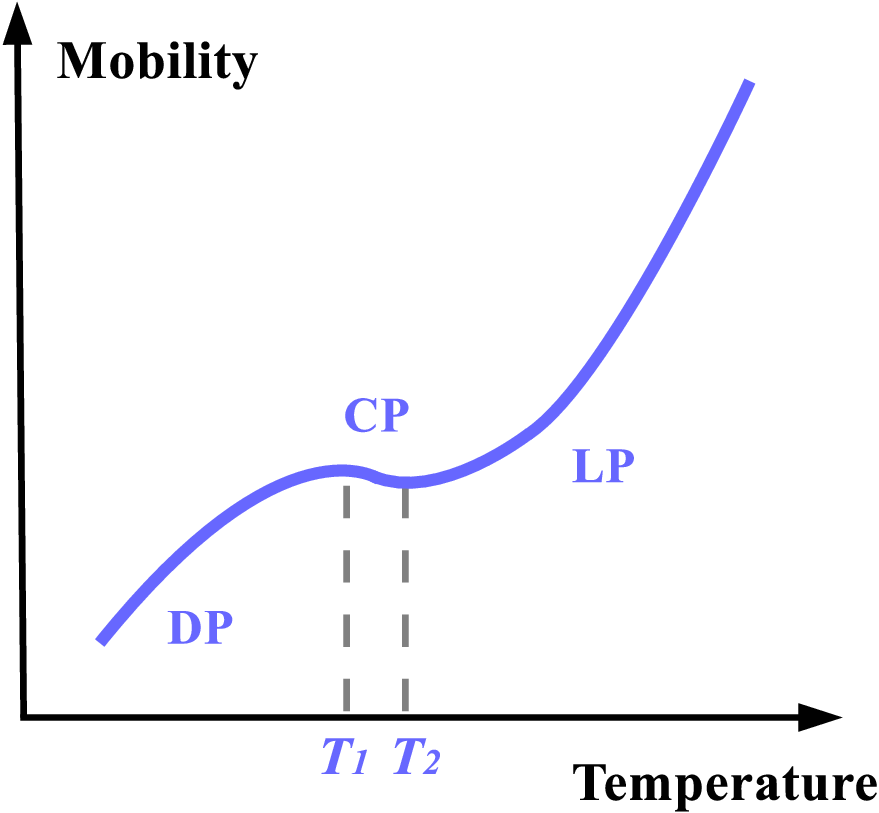}
\caption{Schematic for the mobility-temperature relationship and the category of the localized phase (LP), critical phase (CP) and dispersive phase (DP).}\label{Fig7}
\end{figure}

In common sense of organic semiconductors, the mechanism of charge transport normally refers to a single electron and shifts between bandlike and hopping \cite{Transport1,Transport2}. As sketched in Fig.~\ref{Fig7}, there are three regimes of the $\mu-T$ relationship \cite{Negative1,Negative2,Negative3}. When the temperature is lower than $T_1$ or higher than $T_2$, namely at low or high temperature regime, the mobility increases with different slopes following temperature increasing. In the intermediate regime, the relationship becomes weakly dependent or even negative indicating a completely different mechanism dominates. In field-effect transistors of PAHs, $T_1$ is roughly 200K and $T_2$ is 250K \cite{Negative2}. This novel phenomenon thus inspires the discussion of the bandlike-hopping duality. The bandlike transport normally refers to the coherent motion of electron wavepackets. The thermal fluctuation breaks the quantum coherence so that the $\mu-T$ relationship could be negative \cite{Troisi2}. The suffix ``like" indicates it is not fully band transport as that in the inorganic crystals, since the organic semiconductors are usually amorphous and the electrons are not fully delocalized as stated. The hopping mechanism suggests the electron moves as a classical particle which stochastically hops among the nearest molecules. The thermal fluctuation facilitates the hopping mechanism by either providing energy to cross the potential barriers or eventually pulling the molecules closer \cite{Yao12}.

An experienced experimenter would say that, although one might frequently observe the negative $\mu-T$ relationship, it is quite uncertain. The same material with the same synthesis approach gives rise to completely different transition temperature. In other words, this is a structure-sensitive effect. A viewpoint on the basis of a single electron in a static disorder lattice can not sufficiently explain this exotic phenomenon. Our CT fracton model may help comprehend it. When an additional electron is injected into the D-A alternating structure, it is immobile due to the strong Coulomb attraction, so there is not a usual hopping term (kinetic energy) in the Hamiltonian. The motion of the electron must be enabled by changing the states of phonon modes, that is, there must be vibronic couplings in the Hamiltonian which enforces the electron move. This viewpoint is also similar with that in the fracton polaron model \cite{Fracton0}.

In this new concept, we may also have a more sophisticated scenario to describe the novel $\mu-T$ relationship as depicted in Fig.~\ref{Fig7}. Below the temperature of $T_1$, the phonon modes around 100cm$^{-1}$ are activated which are from secondary intermolecular vibrations. These off-diagonal vibronic couplings act as nonlocal random forces to drive the electron move, and the higher the temperature is, the stronger the force is. As a result, at low temperature regime, it is a nonadiabatic entropy-increasing process, and the phonons play a mediate role in the heat exchange. Above the temperature of $T_2$, the phonon modes larger than 500cm$^{-1}$ are activated which are from primary intramolecular vibrations. These diagonal vibronic couplings induce the localization of electrons, and the correlation between two molecules is quenched by the high-frequency vibrations such that the entropy decreases \cite{Yao12}. Afterward, the normal diffusion of electrons via the incoherent hopping among molecules takes place. Of the most interests is the abnormal regime between $T_1$ and $T_2$, in which both the intra- and inter-molecular vibrations are comparable with each other and then induce aggregate phonons. This is the regime that the critical phase and CT fractons may emerge. The eigen-states are partially localized and the entropy increases very slowly. In this phase, the fracton-like electron moves among different eigen-states via quantum tunnelings solely driven by aggregate phonons (remember no usual hopping terms). It is dominated by a quantum-mechanical probability, and the thermal dissipation does nearly not matter in this process. In a previous work, we call this process as ``coherent hopping", in comparison with the incoherent hopping in the diffusive transport \cite{Yao17}.

To this end, we write down a standard one-dimensional Hamiltonian solely taking both vibronic couplings into account and making the usual hopping terms vanish, i.e.,
\begin{eqnarray}
&&H=\sum_{i,\mu}\lambda_{i,\mu} x_{i,\mu}|i\rangle\langle i|+\sum_{i,\nu}\delta_{\nu} \tilde{x}_{i,\nu}(|i\rangle\langle i+1|+{\rm h.c.})\nonumber\\
&&+\frac{1}{2}\sum_{i}m_i\left[\sum_{\mu}(\dot{x}_{i,\mu}^2+\omega_{\mu}^2x_{i,\mu}^2)+\sum_{\nu}(\dot{\tilde{x}}_{i,\nu}^2+\tilde{\omega}_{\nu}^2\tilde{x}_{i,\nu}^2)\right],\nonumber\\&&
\end{eqnarray}
where $|i\rangle$ is a local state of electron on $i$-th molecule with the local (diagonal) vibronic coupling being $\lambda_{i,\mu}$ and the nonlocal (off-diagonal) coupling being $\delta_{\nu}$ (the same spectral densities with that in the last Subsection); $x_{i,\mu}$ ($\tilde{x}_{i,\nu}$) is the displacement of $i$-th molecule in $\mu$-th local ($\nu$-th nonlocal) vibrational modes with $m_i$ being the mass of the $i$-th molecule and $\omega_{\mu}$ ($\tilde{\omega}_{\mu}$) being the relevant frequency. In order to manifest the characteristics of D-A alternating structure, $\lambda_{i,\mu}$ is set to $\lambda_{\mu}$ on the odd sites and vanishing on even sites, so the system reserves the translational symmetry. We do not consider the usual hopping terms, so the off-diagonal coupling drives the electrons move whose coupling strength $\beta$ is fixed to be 0.1, following our convention in the studies of spin-boson models \cite{Yao14,YaoSR}. The total number of sites is set to 20, and $s$ equals to 0.5 for both couplings. As the transition between localization and dispersion is mainly considered, the finite-size effect and boundary conditions do not matter.

\begin{figure}
\includegraphics[scale=0.7]{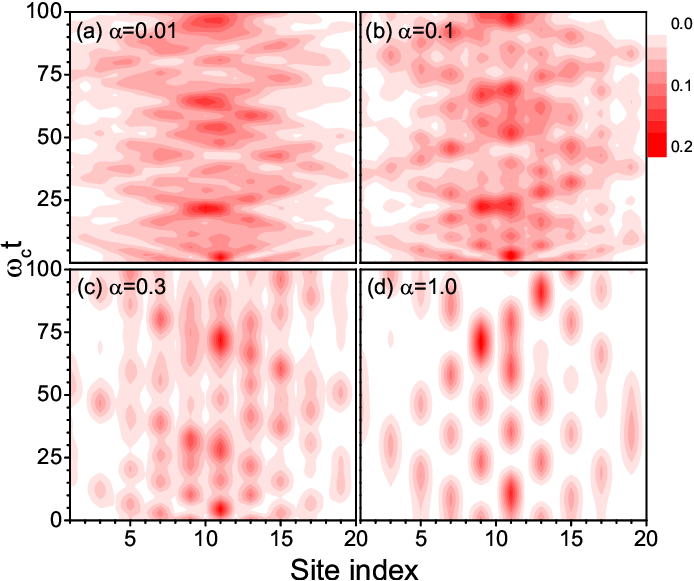}
\caption{Evolution of population distribution on the lattice with two baths and one electron for $\alpha=$ (a) 0.01, (b) 0.1, (c) 0.3 and (d) 1.0.}\label{Fig8}
\end{figure}

The evolution of electron population on the lattice is computed by the adaptive time-dependent density matrix renormalization group algorithm \cite{Yao15,Yao16}, as displayed in Fig.~\ref{Fig8}. Since our CT fracton model mainly corresponds to the parity of CT state and phonons, the initial state is obtained via applying the parity operator $\hat{V}\equiv\exp(i\pi \hat{n}_x)$ onto the ground state, with $\hat{n}_x$ being the number operator of electrons at molecule $x$. Here, we choose $x$ as the central molecule (``$c$") in the lattice. When $\alpha$ is small, the off-diagonal coupling dominates so that the electron freely expands on the lattice. Only finite number of sites are considered so the electrons bounded against the ends and are not able to be fully delocalized. Following $\alpha$ increasing, there emerge visible spatially dispersive states as indicated by the spots, and when $\alpha=1.0$ the electrons are almost localized near the initial site. The competition between diagonal and off-diagonal couplings gives rise to the transition of the mechanisms.

\begin{figure}
\includegraphics[scale=1.1]{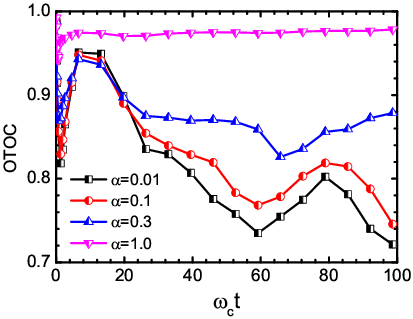}
\caption{Time evolution of OTOC on the lattice with two baths and one electron for $x=1$ as a function of $\alpha$.}\label{Fig9}
\end{figure}

We then calculate the time evolution of the OTOC with the operators defined as $\hat{V}\equiv\exp(i\pi \hat{n}_c)$ and $\hat{W}\equiv\exp(i\pi \hat{n}_{c+x})$. Fig.~\ref{Fig9} shows the results for $x=1$, and results for other $x$'s are quite similar and not shown due to the translational symmetry. It is clear that for $\alpha=0.01$ the OTOC oscillates and decays, stemming from the dephasing among eigen-states. On the other hand, the quantum coherence among different spatial localized states preserves, so the electrons can diffuse. The OTOC will finally decay to zero after long time evolution, which is several orders longer than that of the present calculations \cite{OTOC3,OTOC5}. Following $\alpha$ increases the OTOC increases as well, but still exhibits the dephasing feature. We notice that, the ergodic phase possesses both the dephasing feature and the energy dissipation \cite{MBL0}. Combined with the population distribution as shown in Fig.~\ref{Fig8}, we realize the cases of $\alpha=0.01$ and 0.1 approximately exhibit the dispersive feature while the case of $\alpha=0.3$ manifests the intermediate feature, which can be regarded as in the critical phase. For $\alpha=1$, it keeps at a value close to 1 indicating the electrons are completely localized. It is worth noting that, the value of IPR for $\alpha=1$ can be remarkably larger than one as the sites other than the initial one are also populated, so it fails to reflect the localization length.

\begin{figure}
\includegraphics[scale=0.65]{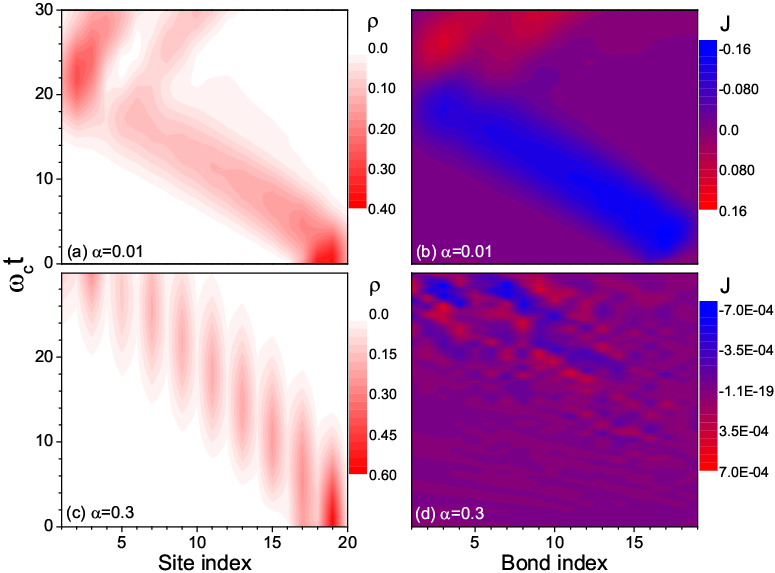}
\caption{Evolution of distribution of population $\rho$ on the lattice with two baths and one electron in the presence of electric field for $V_0=0.001$ and $\alpha=$ (a) 0.01 and (b) 0.3. Evolution of the relevant $J$ (see in the text for definition) are shown in (c) and (d). }\label{Fig10}
\end{figure}

There are several unique issues relating to the localization, such as the absence of the eigen-state thermalization and the dc conductivity \cite{MBL0}. The former one has been discussed in our previous work \cite{AP}, so here we investigate to show the latter. To this end, we set an initial state that the electrons are residing on the right end of the lattice and then add a uniform electric field onto the lattice, i.e. add a potential term $\sum_iiV_0\hat{n}_i$ into the Hamiltonian while time evolving. $V_0$ is set to 0.001, sufficiently small to ensure the system is not remarkably influenced by this term. We calculate both the population distribution and the quantity $J={\rm Im}\langle |i+1\rangle\langle i|-|i\rangle\langle i+1|\rangle$ to figure out the change of current, as displayed in Fig.~\ref{Fig10}. It is found that, the electrons freely move from the right to left for $\alpha=0.01$ in the dispersive phase and behave a hopping-like motion in a very slow manner for $\alpha=0.3$ in the critical phase. The latter behavior is remarkably similar with that of the CT fractons, so we can argue that the fracton model has its available extension in charge transport. The $J$ distribution manifests the same effect, and the values of $J$ for $\alpha=0.3$ are two orders smaller than that for $\alpha=0.01$. More importantly, $J$ for $\alpha=0.3$ exhibit an alternating behavior other than a continuous spatial distribution indicating the absence of dc current. The case for $\alpha=1.0$ is calculated as well (not shown), in which the electrons move much slower than than that in above cases. All these results suggest the dc conductivity is fairly small but the ac conductivity is moderate in the critical phase, so there is normally a kink in the $\mu-T$ curve as shown in Fig.~\ref{Fig7}.

Before ending this Section, let us briefly discuss the measurements of mobility. In the traditional semiconductor physics, the mobility and the mean velocity of charge carriers are dependent on the scattering rate of impurities and phonons \cite{Semi}. The former gives rise to a positive temperature coefficient while the latter refers to negative coefficient. Most techniques of mobility measurements such as the time-of-flight (TOF) approach are based upon the random walk of injected carriers \cite{TOF1}. If the semiconductor is dominated by purely incoherent hopping of localized electrons, the Einstein's relation between mobility and diffusion coefficient is available and the TOF still properly works. The Einstein's relation is applicable in the ergodic phase, and the critical phase do not explicitly fit for TOF \cite{TOF2}. Considering the mixture of coherent and incoherent hopping mechanisms, one can also divide the mobility into coherent and incoherent components, respectively \cite{MobiCohe}. In our opinion, therefore, the mobility measurements should be combined with the time-resolved microwave or terahertz conductivity to determine the component from coherent hopping induced by aggregate phonons as well as the ESR to quantify the feature of the charge carriers.

\subsection{Singlet fission}

Spin is an important degree of freedom to enable some kinds of quantum effects. In organic semiconductors, both polarons and triplet excitons carry non-zero spin. So far, the organic spintronics mainly refers to the excited states, such as inter-system crossing, triplet-triplet annihilation, triplet-polaron quench, and so on. There are very few researches on the magnetism of the ground state, which may demand for ordered structure in macroscopic scale. In mesoscopic level as stated, the singlet fission is an interesting issue. This is because the generated triplet excitons possess much lower energy than that of the singlet Frenkel exciton, so they act more likely low-energy excitations from ground state. Actually, if we forget D's LUMO and A's HOMO which do not matter in our toy model, the triplet excitons have got almost the same characteristics with the CT excitons. So we can reasonably guess that the CT fracton mechanism can be applicable in singlet fission. In order to avoid confusion with the commonly-presented intermediate CT state in singlet fission, however, we still use the conventional terminology of triplet excitons in the following discussion.

The singlet fission that a singlet exciton is split into two triplet excitons turns out to be an appealing effect to facilitate the efficiency of photovoltaics. However, a usual doubt on the statement that the singlet fission can fairly enhance the power conversion efficiency is, although one singlet is split into two triplets and subsequently two electron-hole pairs might be formed to increase the short-circuit current, the energy of these electron-hole pairs are lost in the singlet fission and thus the open-circuit voltage should be decreased. As a result, the efficiency can not be significantly enhanced after all.

In order to comprehend this puzzle, it is again useful to employ the framework of quantum heat engine model. Let us assume the dissociation of each triplet exciton consumes the same energy with that discussed in the photovoltaics, giving the fact that the entanglement entropy which is quenched in the exciton dissociation is insensitive to the energy of excitons. Since the relaxation processes are almost absent in singlet fission materials, the singlet energy ($Q_1$) must be doubled to keep the quantum efficiency $\eta$ unchanged. We notice that, the singlet fission materials exhibit higher external quantum efficiency for larger photon energy \cite{EQE}, so they are indeed more suitable for the ultraviolet regimes. In the practical application of photovoltaics, one should then add additional absorbers into the device to absorb the low-energy photons. In ideal cases, the maximum efficiency will increase from 45.7$\%$ to $47.7\%$ by inserting an additional layer \cite{SQ1,SQ2}. On the other hand, the transition from triplet pair state (TT state) to the uncorrelated two triplets (T+T state) is also an entropy-increasing process which may further lower the efficiency. Therefore, it is necessary to decrease the energy dissipation in TT state dissociation as much as possible, and we have to carefully investigate the dissociation process of TT state.

A big question now arises: What makes the TT state dissociate? In absence of the dissipated environment, the two triplet excitons are consistently entangled with each other, no matter how long they separate in the real space. Therefore, the dissociation means the loss of coherence between two triplet excitons, analogous to the CT state dissociation. The triplet exciton is of charge neutrality and large binding energy so in terms of charge degree of freedom the exciton-phonon couplings should be relatively weak. Meanwhile, the organic materials are basically composed of light elements with negligible spin-orbital couplings, so the spin coherence is also of long lifetime. Fortunately, there are still many factors related to the spin degree of freedom, such as the hyperfine interaction of hydrogen atoms \cite{Hyper}, the hybridization of $\sigma$ and $\pi$ orbitals due to out-of-plane vibrations \cite{Yu}, the other triplet excitons \cite{Gillin}, the kinetically blocked radicals \cite{FLi} and also the rotation with different chirality of polarized molecules. All these factors can be grouped into a local spin-related bath for the electrons, holes and triplet excitons, which may influence the decoherence and localization of TT state.

The intermolecular distance may also serve as a key factor in the TT dissociation. The distance must be sufficiently short such that the electron transfer integral is large enough to induce the conversion of local singlet exciton and the TT state. The short distance will however enhance the spin-spin exchange interaction between triplets and make the TT dissociation unfavorable. This conflict heavily limits the practical application of singlet fission in photovoltaics. Therefore, the spin-spin interaction which is the main interaction for delocalization of spin should be carefully investigated as well.

To this end, let us write down a Hamiltonian for two pairs of electron and hole taking the diagonal bosonic environment into account. That is
\begin{eqnarray}
&&H=-2J\hat{\mathbf{S}}_1\hat{\mathbf{S}}_2+D\sum_{i=1,2}\hat{S}_{i,z}^2+E\sum_{i=1,2}(\hat{S}_{i,x}^2-\hat{S}_{i,y}^2)\nonumber\\&&+\sum_{i,\nu}\lambda_{i,\nu} \hat{S}_{i,z}(\hat{b}^{\dag}_{i,\nu}+\hat{b}_{i,\nu})+\sum_{i,\nu}\omega_{i,\nu}\hat{b}^{\dag}_{i,\nu}\hat{b}_{i,\nu}.
\end{eqnarray}
where $\hat{\mathbf{S}}_i$ is the spin operator for the $i$-th triplet, $J$ is the spin-spin interaction between two triplets, $D$ and $E$ denote the interaction strengths with respect to the molecular field, and the environmental part takes the same formula with those in the last two Subsection. Herein, the first line is nothing but the widely-adopted Merrifield model considering the dipole-dipole interaction \cite{Merri1,Merri2,Merri3}, which initially stimulated us to consider the triplet excitons as CT fractons. The parameters are set as usual cases, namely $D=0.15{\rm cm}^{-1}$ and $E=0.1{\rm cm}^{-1}$.

Different from charge degree of freedom which can off-diagonally couple to the intermolecular vibrations, the spin-related environment is merely diagonal such that the spin-environment coupling itself does not change the population on each spin state \cite{Siwei}. There are two parameters influencing the change of local populations: $J$ and $E$. As we are focusing on the dissociation of triplet pairs, the influence of $J$ is mainly considered which is sensitive to the distance between triplets as stated. The effect of the spin-spin interaction is the conversion of the states $|+1,-1\rangle$, $|-1,+1\rangle$ and the state $|0,0\rangle$, with $\pm 1$ and 0 being the common notations of the relevant state for the 1st and 2nd triplet in order. Now, let us assume the singlet state is initially conversed to the TT states $|+1,-1\rangle$ and $|-1,+1\rangle$, and they entangle with relevant environmental states. The quantum state can be expressed as
\begin{eqnarray}
a|+1,-1,\mathcal{E}_{+-}\rangle+b|-1,+1,\mathcal{E}_{-+}\rangle,
\end{eqnarray}
where $\mathcal{E}$ denotes the relevant environmental states, and $a$, $b$ the combination coefficients. In order to make the TT state dissociate, each triplet must turn to entangle with their own local environmental state. That is, the quantum state have to change to
\begin{eqnarray}
|\sigma_1,\mathcal{E}_{1}\rangle\otimes|\sigma_2,\mathcal{E}_{2}\rangle,
\end{eqnarray}
where $\sigma_i$ is a linearly combined spin state locally for $i$-th triplet. The larger change of environmental state is, the easier the decoherence takes place, so the entropy change heavily depends on the change of the environmental states. Interestingly, however, the environmental states are hardly changed and thus both $|+1,-1\rangle$ and $|-1,+1\rangle$ are not easy to be dissociated. This is because, during the time evolution both $|+1,-1\rangle$ and $|-1,+1\rangle$ are firstly conversed to the state $|0,0\rangle$, and $|0,0\rangle$ does not couple to the environment in terms of the present Hamiltonian. As a result, if we want to increase the yield of free triplet excitons, it is essential to decrease the yield of $|+1,-1\rangle$ and $|-1,+1\rangle$ states during the singlet-TT conversion. This can be done by, e.g., exerting an external magnetic field to open energy gaps among +1, 0 and -1 states \cite{Merri3}.

\begin{figure}
\includegraphics[scale=0.65]{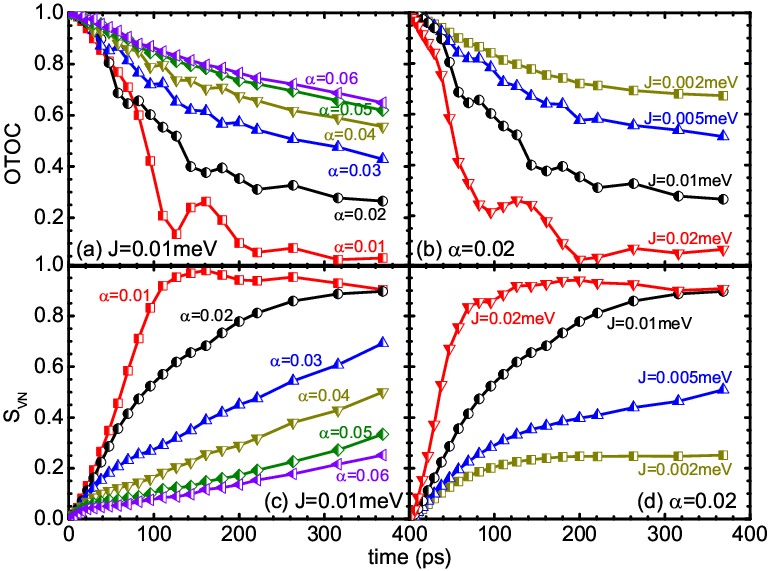}
\caption{Time evolution of OTOC in the Merrifield model with spin-related environment for singlet fission as a function of (a) $\alpha$ for $J=0.01{\rm meV}$ and (b) $J$ for $\alpha=0.02$. The relevant evolution of von Neumann entropy $S_{\rm VN}$ is shown in (c) and (d).}\label{Fig11}
\end{figure}

The $|0,0\rangle$ state is easier to be dissociated since it firstly converses to either $|+1,-1\rangle$ or $|-1,+1\rangle$ state during the time evolution. We then calculate the time evolution with the initial state being that, on each molecule the electron and hole form a state $\mid\uparrow\downarrow\rangle$, namely the combination of the singlet and the 0 component of triplet. Both the OTOC and von Neumann entropy $S_{\rm VN}$ are calculated for the two electrons of the excitons on the two molecules, as displayed in Fig.~\ref{Fig11}. It is found that, with $J=0.02$meV and $\alpha\leq0.02$ the system is delocalized, i.e. the correlation between two triplets is strong. In this situation the TT state is hard to be dissociated. When $J$ is between 0.005 and 0.01meV, the behavior of OTOC exhibits the oscillation feature which may make the dissociation much easier. For $J=0.002$meV it becomes localized. Therefore, $J=$0.01meV (or equivalently 90mT) and $\alpha=0.02$ emerge to be critical values for efficient dissociation of TT state. We also calculate $S_{\rm VN}$ for the local electrons. The basic feature of $S_{\rm VN}$ is similar with that of OTOC, but some of the details such as the oscillations become weaker exhibiting the advantages of using OTOC. Interestingly, we find very similar behavior of $S_{\rm VN}$ with that in the many-body model with electron-electron interactions \cite{MBL4}, that is, the $S_{\rm VN}$ firstly saturates and then increases slowly.

In the usual study of Shockley-Queisser's detailed balance model for the singlet fission, the spin degree of freedom is not taken into account \cite{SQ2}. People mainly focused on the energy profile of singlet, triplet and CT state and dedicated to synthesize exergonic singlet fission materials. Whereas, the TT state must be firstly dissociated into free triplets before they are able to provide two pairs of free electron and hole. As calculated, the entropy increase and thus the free energy decrease in the TT dissociation are considerable. Different from that for charge degree of freedom, the energy scale for the spin systems is normally 2-4 orders smaller than the thermal fluctuation energy at room temperature. It is thus not able to compensate the energy dissipation in TT dissociation through designing the appropriate energy profiles of different spin states. Fortunately, the singlet fission is an issue of dynamics, not only energetics. Properly dealing with the quantum coherence and entanglement will greatly facilitate the improvement of efficiency. To be more specific, due to the difference of energy scales, the spin dynamics is much slower than the charge dynamics, so the quench of entanglement at proper time point may largely change the yield of free triplets. Consequently, a rational design of high-efficiency singlet fission materials needs very detailed dynamics simulations in addition to the energetic computations.

\section{Summary and outlook}

In summary, the robustness of quantum effects in organic semiconductors under ambient circumstance has been investigated with the theoretical framework of CT excitons and aggregate phonons. D-A alternating structure is the first important factor that gives rise to the liquid of CT states. Excitations in the CT liquid, so-called CT excitons, can move in an undissipative manner in absence of phonons. Phonons are normally regarded to be negative ingredients of quantum effects, but we propose a mechanism that if a number of phonon modes are quenched in aggregate except those resonant ones in multiple molecules, the immobile CT excitons become mobile triggered by thermal fluctuations. This so-called CT fracton model enables the positive influence of aggregate phonons. On the basis of these phenomenological toy models, the hierarchy of quantum effects in various materials such as PAHs, metal-organic complexes and CT salts are discussed in details. The processes of photoelectric conversion, charge transport and singlet fission are analyzed as well. The quantum heat engine model is important and often mentioned in the field of organic solar cells, since if one wants to improve the power conversion efficiency in an ideal case or even break the Shockley-Queisser's limit, the consideration of the quantum effects in the heat engine model is remarkable. We then consider two kinds of competing interactions --- one refers to the localization and the other to delocalization. Frenkel-CT mixed model, two-bath lattice model and the Merrifield model for singlet fission are studied by the adaptive time-dependent density matrix renormalization group algorithm and the time evolution of OTOC is respectively calculated. The three models we studied involve charge, spin and vibronic interactions, covering the majority cases of the theories in the field. Therefore, our research makes a step forward to establish a unified quantum dynamic theory relating to charge, exciton, spin and phonons in organic semiconductors.

The statistical mechanics relies on the basic postulate of the ergodic hypothesis, which straightforwardly leads to the principle of entropy increase and free energy decrease. As discussed, however, the organic semiconductors are rarely in the ergodic phase, and the disorders and interactions give rise to complicated transition between localization and dispersion. It is therefore insufficient to analyze the issues in organic semiconductors simply with the argument from the energetics, such as the electron naturally hop from the higher-energy level to the lower one. The dynamic information turns out to be essential. Especially, the time of decoherence and disentanglement in the dynamics, which is dependent of the competing vibronic couplings, plays a significant role in affecting the bandlike and hopping mechanism, the ultrafast long-range charge transfer, the triplet pair dissociation, and other relevant processes. As a consequence, the proper understanding of quantum effects may tell us the explicit working mechanisms of the realistic devices.

Throughout this work, we quantitatively determined parameter regimes that CT fractons matter but merely investigated the dynamics of entanglement entropy in a qualitative manner. One would intuitively consider that how the change of entropy in realistic devices can be explicitly quantified. Commonly speaking, both the von Neumann entropy and OTOC are defined between 0 and 1. We have to properly calibrate them to fit for the standard statistical mechanics and the relevant experiments. This is not easy in a general manner but we can first try to bridge the entanglement entropy and the entropy in organic solar cells which will be the next subject.

\section*{Acknowledgment}

The author gratefully acknowledges support from the National Natural Science Foundation of China (Grant Nos.~91833305, 11974118), Key Research and
Development Project of Guangdong Province (Grant No.~2020B0303300001), Guangdong-Hong Kong-Macao Joint Laboratory of Optoelectronic and Magnetic Functional Materials program (No.~2019B121205002) and Fundamental Research Funds for the Central Universities (Grant No.~2019ZD51). We thank Prof. Haibo Ma and Yang Zhao for fruitful discussions.

\end{document}